\documentclass[twocolumn]{aastex61}
\usepackage{amsmath}
\usepackage{mathtools}

\begin{document}

\title{On the ultra-relativistic Prompt Emission (UPE), the Hard and  Soft X-ray Flares, and the extended thermal emission (ETE) in GRB 151027A}

\author{R.~Ruffini}
\affiliation{ICRA and Dipartimento di Fisica, Sapienza Universit\`a di Roma, P.le Aldo Moro 5, 00185 Rome, Italy}
\affiliation{ICRANet, P.zza della Repubblica 10, 65122 Pescara, Italy.}
\affiliation{Universit\'e de Nice Sophia Antipolis, CEDEX 2, Grand Ch\^{a}teau Parc Valrose, Nice, France}
\affiliation{ICRANet-Rio, Centro Brasileiro de Pesquisas F\'isicas, Rua Dr. Xavier Sigaud 150, 22290--180 Rio de Janeiro, Brazil. \href{mailto:ruffini@icra.it}{ruffini@icra.it}}

\author{L.~Becerra}
\affiliation{ICRA and Dipartimento di Fisica, Sapienza Universit\`a di Roma, P.le Aldo Moro 5, 00185 Rome, Italy}
\affiliation{ICRANet, P.zza della Repubblica 10, 65122 Pescara, Italy.}

\author{C.~L.~Bianco}
\affiliation{ICRA and Dipartimento di Fisica, Sapienza Universit\`a di Roma, P.le Aldo Moro 5, 00185 Rome, Italy}
\affiliation{ICRANet, P.zza della Repubblica 10, 65122 Pescara, Italy.}

\author{Y.~C.~Chen}
\affiliation{ICRA and Dipartimento di Fisica, Sapienza Universit\`a di Roma, P.le Aldo Moro 5, 00185 Rome, Italy}
\affiliation{ICRANet, P.zza della Repubblica 10, 65122 Pescara, Italy.}

\author{M.~Karlica}
\author{M.~Kova\v{c}evi\'{c}}
\affiliation{ICRA and Dipartimento di Fisica, Sapienza Universit\`a di Roma, P.le Aldo Moro 5, 00185 Rome, Italy}
\affiliation{ICRANet, P.zza della Repubblica 10, 65122 Pescara, Italy.}
\affiliation{Universit\'e de Nice Sophia Antipolis, CEDEX 2, Grand Ch\^{a}teau Parc Valrose, Nice, France}
\author{J.~D.~Melon~Fuksman}
\author{R.~Moradi}
\author{M.~Muccino}
\author{G.~B.~Pisani}
\author{D.~Primorac}
\affiliation{ICRA and Dipartimento di Fisica, Sapienza Universit\`a di Roma, P.le Aldo Moro 5, 00185 Rome, Italy}
\affiliation{ICRANet, P.zza della Repubblica 10, 65122 Pescara, Italy.}

\author{J.~A.~Rueda}
\affiliation{ICRA and Dipartimento di Fisica, Sapienza Universit\`a di Roma, P.le Aldo Moro 5, 00185 Rome, Italy}
\affiliation{ICRANet, P.zza della Repubblica 10, 65122 Pescara, Italy.}
\affiliation{ICRANet-Rio, Centro Brasileiro de Pesquisas F\'isicas, Rua Dr. Xavier Sigaud 150, 22290--180 Rio de Janeiro, Brazil. \href{mailto:ruffini@icra.it}{ruffini@icra.it}}

\author{G.~V.~Vereshchagin}
\affiliation{ICRA and Dipartimento di Fisica, Sapienza Universit\`a di Roma, P.le Aldo Moro 5, 00185 Rome, Italy}
\affiliation{ICRANet, P.zza della Repubblica 10, 65122 Pescara, Italy.}

\author{Y.~Wang}
\affiliation{ICRA and Dipartimento di Fisica, Sapienza Universit\`a di Roma, P.le Aldo Moro 5, 00185 Rome, Italy}
\affiliation{ICRANet, P.zza della Repubblica 10, 65122 Pescara, Italy.}

\author{S.~S.~Xue}
\affiliation{ICRA and Dipartimento di Fisica, Sapienza Universit\`a di Roma, P.le Aldo Moro 5, 00185 Rome, Italy}
\affiliation{ICRANet, P.zza della Repubblica 10, 65122 Pescara, Italy.}

\begin{abstract}
We analyze GRB 151027A within the binary-driven hypernova (BdHN) approach, with progenitor a carbon-oxygen core on the verge of a supernova (SN) explosion and a binary companion neutron star (NS). The hypercritical accretion of the SN ejecta onto the NS leads to its gravitational collapse into a black hole (BH), to the emission of the GRB and to a copious $e^+e^-$ plasma. The impact of this $e^+e^-$ plasma on the SN ejecta explains the early soft X-ray flare observed in long GRBs. We here apply this approach to the UPE and to the hard X-ray flares. We use GRB 151027A as a prototype. From the time-integrated and the time-resolved analysis we identify a double component in the UPE and confirm its ultra-relativistic nature. We confirm the mildly-relativistic nature of the soft X-ray flare, of the hard X-ray flare and of the ETE.  We show that the ETE identifies the transition from a SN to the HN. We then address the theoretical justification of these observations by integrating the hydrodynamical propagation equations of the $e^+ e^-$ into the SN ejecta, the latter independently obtained from 3D smoothed-particle-hydrodynamics simulations. We conclude that the UPE, the hard X-ray flare and the soft X-ray flare do not form a causally connected sequence: Within our model they are the manifestation of \textbf{the same} physical process of the BH formation as seen through different viewing angles, implied by the morphology and the $\sim 300$~s rotation period of the HN ejecta.
\end{abstract}


\section{Introduction}\label{sec1}

Gamma-ray bursts (GRBs) are traditionally classified in short GRBs, with a total duration $\lesssim 2$~s, and long GRBs, lasting $\gtrsim 2$~s \citep{1981Ap&SS..80....3M,Dezalay1992,Klebesadel1992,Kouveliotou1993,Tavani1998}. Large majority of long bursts is spatially correlated with bright star-forming regions in their host galaxies \citep{Fruchter2006,Svensson2010}. {For this reason the long GRBs have been traditionally associated with  the collapse of the core of a single  massive star to a black hole (BH),  surrounded by a thick massive accretion disk: the \textit{collapsar} \citep{1993ApJ...405..273W,1998ApJ...494L..45P,1999ApJ...524..262M,2004RvMP...76.1143P,2013ApJ...764..179B}. In this traditional picture the GRB dynamics follows the ``fireball'' model, which assumes the existence of a single ultra-relativistic collimated jet \citep[see e.g.][]{1976PhFl...19.1130B,1990ApJ...365L..55S,1993MNRAS.263..861P,1993ApJ...415..181M,1994ApJ...424L.131M}. The structures of long GRBs were described either by internal or external shocks \citep[see][]{1992MNRAS.258P..41R,1994ApJ...430L..93R}. The emission processes were linked to the occurrence of synchrotron and/or inverse-Compton radiation coming from the single ultrarelativistic jetted structure, characterized by Lorentz factors $\Gamma \sim 10^2$--$10^3$.}

{Such a \textit{collapsar} model does not address some observational facts: 1) most massive stars are found in binary systems \citep{Smith2014}, 2) most type Ib/c SNe occur in binary systems \citep{Smith2011} and 3) the SNe associated to long GRBs are indeed of type Ib/c \citep{DellaValle2011}. These facts motivated us to develop the binary-driven hypernova (BdHN) model.}

Recently we have found evidence for multiple components in long GRB emissions, indicating the presence of a sequence of astrophysical processes \citep{2012A&A...543A..10I,2012A&A...538A..58P}, which have led to formulate in precise terms the sequence of events in the Induced Gravitational Collapse (IGC) paradigm \citep{Ruffini2001c,Ruffini2007,Rueda2012,Fryer2014} making explicit the role of binary systems as progenitors of the long GRBs.

Within the IGC scenario the long bursts originate in tight binary systems composed of a carbon-oxygen core (CO$_{\rm core}$) undergoing a SN explosion and a companion neutron star (NS) \citep{Becerra,2016ApJ...833..107B,2018arXiv180304356B}. The SN explosion triggers a hypercritical accretion process onto the companion NS: photons are trapped in the infalling material and the gravitational energy gained by accretion is carried out through an efficient neutrino emission \citep{Zeldovich1972,RRWilson1973,Fryer2014}. Depending on the CO$_{\rm core}$-NS binary separation/period two outcomes may occur. For widely separated ($a\gtrsim10^{11}$~cm) CO$_{\rm core}$-NS binaries, the hypercritical accretion rate is $<10^{-2}~M_\odot$~s$^{-1}$ and it is insufficient to induce gravitational collapse of the NS to a BH. Instead, the NS just increases its mass becoming a massive NS. This process leads to the emission of the so-called X-ray flashes (XRFs) with a typical X-ray emission $\lesssim 10^{52}$~erg. 

For more tightly bound ($a\lesssim10^{11}$~cm) CO$_{\rm core}$-NS binaries the hypercritical accretion rate of the SN ejecta can be as large as $\gtrsim10^{-2}$--$10^{-1}~M_\odot$~s$^{-1}$, leading the companion NS to collapse to a BH.
This process leads to the occurrence of {the BdHN} which exhibits a more complex structure than XRFs and an emission $\gtrsim 10^{52}$~erg \citep{2016ApJ...832..136R}.

The opportunity of introducing the BdHN model, based on binary progenitors, exhibiting  a large number of new physical process and admitting a theoretical treatment by detailed equations whose corresponding solutions are in agreement with the observations, has been presented in a large  number of publications, recently summarized in \citet{2018ApJ...852...53R}.
There we performed an extensive analysis using 421 BdHN all with measured redshift, observed till the end of 2016, and described in their cosmological rest frame \citep{2016ApJ...833..159P}.

The large variety of spectra and light curves has allowed the introduction of seven different GRBs subclasses, see e.g. \citet{2016ApJ...832..136R} and \citet{2016arXiv160203545R}.

We recalled that since 2001 we fit the Ultra-relativistic Prompt emission (UPE) light curve and spectra solving the equations of the dynamics of the $e^+e^-$ baryon plasma and of {its slowing down due to the interaction} with the circumburst medium \citep[CBM, see e.g.][]{1999A&A...350..334R,2002ApJ...581L..19R,2000A&A...359..855R}.
This treatment allows to evaluate  the ultra-relativistic gamma factor of the UPE, exhibited in hundreds of short and long GRBs. Some underluminous GRBs may well have a non-ultrarelativistic prompt emission (Rueda et al., in preparation).

{ 
Attention was then directed to examine the Flare-Plateau-Aftergolw phase (FPA) following the UPE.}

{
We identified among the BdHNe \textit{\textbf{all}} the ones with soft X-ray flare in the $0.3$--$10$~keV rest-frame energy range in the FPA phase.
In view of the excellent data and complete light-curves we could identify in them a thermal component, see Fig.~32 and Table.~7 in \citet{2018ApJ...852...53R},  essential in measuring the mildly relativistic} expansion velocity of $v = c \beta \sim 0.8 c$, see section 9 in \citet{2018ApJ...852...53R}.

{
In addition we then followed, through an hydrodynamical description, the propagation {and the slowing down} inside the SN ejecta of the $e^+e^-$ plasma generated in the BH formation, in order to explain the mildly relativistic nature of the soft X-ray flares expansion velocity, see section 10 in \citet{2018ApJ...852...53R}.} 

{
Obviously these considerations cannot be repeated here.}

{We only recall a few points of the conclusions of \citet{2018ApJ...852...53R}, e.g. a) The data of the soft X-ray flare have determined its mildly relativistic expansion velocity already $\sim 100$~s after the UPE, in contrast to the traditional approach; b) the role of the interaction of the $e^+ e^-$ GRB emission in SN ejecta in order to explain the astrophysical origin of soft X-ray flare; c) the determination of the density profile of the SN ejecta derived from the simulation of the IGC paradigm.}

{
In this article we apply our model to study a multiple component in the UPE phase observed in the range of $10$--$1000$ keV as well as the Hard X-ray Flares observed in the range of $0.3$--$150$ keV, the extended-thermal-emission (ETE), and finally the soft X-ray flare observed in the range of $0.3$--$10$ keV using GRB151027A as a prototype. The aim is to identify the crucial role of the SN and of its binary NS companion in the BdHN model, to analyze the interaction of the $e^+ e^-$ plasma generating the GRB with the SN ejecta via 3D simulations, and to compare and contrast the observational support of the BdHN model with the other traditional approaches. For facilitating the reader we have made a special effort in giving reference to the current works, in indicating new developments and their observational verifications, and finally in giving references for the technical details in the text.}

{
In section~\ref{sec:progress_of_bdhn} we outline the new results motivating our paper:
1) Three thermal emissions processes in GRBs compared and contrasted. Particularly relevant for our article is the relativistic treatment relating the velocity of expansion of the hard X-ray flare, of the soft X-ray flare, and of the ETE to the observed fluxes and temperatures.
2) The 3D simulations of the hypercritical accretion in a BdHN, essential for obtaining the density profiles of the SN ejecta recently submitted for publication in \citet{2018arXiv180304356B}.
3) The generalization of the space-time representation of the BdHN. 
These are some useful conceptual tools needed to create a viable GRB model.
}

{
In section~\ref{sec2} we refer to  GRB 151027A as a prototype example of high quality data, enabling the detailed time-resolved analysis for the UPE phase, with its thermal component, as well as the first high quality data for studying the hard X-ray flare, and especially the clear evolution of the ETE. We perform the time-integrated analysis for the UPE, we further analyze the two ultra-relativistic gamma-ray spikes in the UPE, and apply to the first spike the fireshell model and identify the P-GRB, the baryon load $B=(1.92\pm0.35)\times10^{-3}$ and an average CBM density of $(7.46\pm1.2)$~cm$^{-3}$ which are consistent with our numerical simulation presented in section~\ref{sec5}. We determine {an initial Lorentz factor of the UPE} $\Gamma_0=503\pm76$ confirming the clearly observed ultra-relativistic nature of the UPE.
}

{
In section~\ref{sec3} we perform the time-resolved analysis for the hard X-ray flare and the soft X-ray flare, comparing and contrasting our results with the ones in the literature by \cite{2017A&A...598A..23N}. The hard X-ray flare is divided into 8 time intervals and we find a high significant thermal component existing in all time intervals (see Fig.~\ref{fig4}). We report the results of our time-resolved spectral analysis in the first five columns of Table~\ref{tab1}. Using the best-fit model for non-thermal component in the time interval $95$--$130$~s we determine a Lorentz factor $\Gamma = 3.28 \pm 0.84$ for the hard X-ray flare duration. The soft X-ray flare is analyzed in 4 time intervals, in which spectra are best fitted by a single power-law.
}

{
In section~\ref{sec4} we turn to the thermal component evolving across the hard X-ray flare by adopting the description in the GRB laboratory frame. Following our recent works \citep{2018ApJ...852...53R}, we determine the expansion velocity evidencing the transition from an initial velocity $\approx0.38~c$ and increasing up to $0.98~c$ in the late part, see column 6 of Table~\ref{tab1}. This is the first relativistic treatment of the hard X-ray flare and its associated thermal emission clearly evidencing the transition from a SN to an HN, first identified in GRB 151027A. We compare and contrast our results with the current ones in the literature.
}

{
In section~\ref{sec5} we proceed to the hard X-ray flare and the soft X-ray flare theoretical explanation from the analysis of the $e^+e^-$ plasma propagating {and slowing down} within the SN ejecta. The simulated velocity and radius of the hard X-ray flare and the soft X-ray flare are consistent with the observations. We visualize all these results by direct comparison of the observational data by Swift, INTEGRAL, Fermi and Agile, in addition to the optical observations, with the theoretical understanding of the 3D dynamics of the SN recently jointly performed by our group in collaboration with the Los Alamos National Laboratory \citep{2018arXiv180304356B}. This visualization is particularly helpful in order to appreciate the novel results made possible by the BdHN paradigm and also by allowing the visualization of a phenomena observed today but occurred 10 billion light years away in our past light cone. The impact of the $e^+e^-$ plasma on the entire SN ejecta gives origin to the thermal emission from the external surface of the SN ejecta and, equally, we can therefore conclude that the UPE, the hard X-ray flare and the soft X-ray flare are not a causally connected sequence (see Figs.~\ref{fig:Carlo2}, \ref{fig:model1}, \ref{fig:gamma_ray_flare}, \ref{fig:x_ray_flare} and Tab.~\ref{tab1}): Within our model they are the manifestation of the same physical process of the BH formation as seen through different viewing angles, implied by the morphology and by the $\sim 300$~s rotation period of the HN ejecta.
}

{
In section~\ref{sec6} we proceed to the summary, discussion and conclusions:
\begin{itemize}
\item In the summary we have recalled the derived Lorentz gamma factor and the detailed time resolved analysis of the light curves and spectra of UPE, hard X-ray flare, ETE and soft X-ray flare. We mention a double spike structure in the UPE and in the FPA, which promises to be directly linked to the process of the BH formation. We have equally recalled our relativistic treatment of the ETE, which has allowed to observe for the first time the transition of a SN into a HN: the main result of this paper.
\item In the discussions we have recalled, using specific examples in this article, that our data analysis is performed within a consistent relativistic field-theoretical treatment. In order to be astrophysically significant, it needs the identification of the observed astrophysical components, including: the binary nature of the progenitor system, the presence of a SN component and it needs as well a 3-dimensional simulation of the process of hypercritical accretion in the binary progenitors. We have also recalled the special role of the rotation by which phenomena, traditionally considered different, are actually the same phenomenon as seen from different viewing angles.
\item In the conclusions, looking forward, three main implications follows from the BdHN model which are now open to further scrutiny: 1) only $10$\% of the BdHNe whose line of sight lies in the equatorial plane of the progenitor binary system are actually detectable, in the other $90$\% the UPE is not detectable due to the morphology of the SN ejecta (see Fig.~\ref{fig:cc}) and therefore the \textit{Fermi} and \textit{Swift} instruments are not triggered; 2) the $E_\mathrm{iso}$, traditionally based on a spherically symmetric equivalent emission, has to be replaced by an $E_\mathrm{tot}$ duly taking into account the contributions of the UPE, hard X-ray flare, ETE and soft X-ray flare; 3) when the BdHNe are observed normally to the orbital plane, the GeV emission from the newly formed BH becomes observable and also this additional energy should be accounted for.
\end{itemize}
}

\begin{table}
\centering
\begin{tabular}{lc}
\hline\hline
Extended wording & Acronym \\
\hline
Binary-driven hypernova & BdHN \\
Black hole& BH \\
Carbon-oxygen core& CO$_{\rm core}$ \\
Circumburst medium& CBM \\
Extended thermal emission &ETE\\
Flare-Plateau-Afterglow & FPA \\
Gamma-ray burst& GRB \\
Gamma-ray flash& GRF \\
Induced gravitational collapse & IGC \\
Massive neutron star& MNS \\
Neutron star& NS \\
New neutron star& $\nu$NS \\
Ultra-relativistic prompt  emission & UPE \\
Proper gamma-ray burst & P-GRB \\
Short gamma-ray burst& S-GRB \\
Short gamma-ray flash& S-GRF \\
Supernova& SN \\
Ultrashort gamma-ray burst & U-GRB \\
White dwarf& WD \\
X-ray flash& XRF \\
\hline
\end{tabular}
\caption{{Alphabetic ordered list of the acronyms used in this work.}}
\label{acronyms}
\end{table}

{
We summarize in Table~\ref{acronyms} the list of acronyms introduced in the present paper.}

\section{Recent Progress on BdHNe}
\label{sec:progress_of_bdhn}

{We address three progresses obtained in the last year in the theory of BdHNe: 1) the identification of three different thermal emission processes; 2) the visualization of the IGC paradigm; and 3) an extended space-time diagram of BdHN  with viewing angle in the equatorial plane of the binary progenitors.}

{
One of the first examples of a thermal emission has been identified in the early seconds after the trigger of some long GRBs \citep{Ryde2004,Rydeetal2006,Ryde2009}. This emission has been later identified in the BdHN model with the soft X-ray emission occurring in the photosphere of convective outflows in the hypercritical accretion process from the newly born SN into the NS binary companion. Additional examples have been given in BdHNe \citep{Fryer2014} and in XRFs \citep{2016ApJ...833..107B}. These process are practically Newtonian in character with velocity of expansions of the order of $10^8$--$10^9$~cm~s$^{-1}$ \citep[see e.g.][for the case of GRB 090618]{2012A&A...543A..10I}.
}

{
A second thermal emission process has been identified in the acceleration process of GRBs, when the self-accelerating optically thick $e^+e^-$ plasma reaches transparency and a thermal emission with very high Lorentz factor $\Gamma \sim 10^2$--$10^3$ is observed. This has been computed both in the fireball model \citep{1999PhR...314..575P,2002MNRAS.336.1271D,2007ApJ...664L...1P} and in the fireshell model \citep{RSWX2,2000A&A...359..855R}. The difference consists in the description of the equations of motion of the fireball assumed in the literature and instead explicitly evaluated in the fireshell model from the integration of classical and quantum magnetohydrodynamic process \citep[see also][and references therein]{2007ralc.conf..402R}. The moment of transparency leads to a thermal emission whose relativistic effect have been evaluated leading to the concept of the equitemporal surface \citep[EQTS][]{Bianco2005a}. This derivation has been successfully applied also to short GRBs \citep{2017ApJ...844...83A,2016ApJ...831..178R,2015ApJ...808..190R}, and is here applied in section \ref{sec2} to the UPE.
}

\begin{figure*}[!ht]
\centering
\includegraphics[width=0.95\hsize,clip]{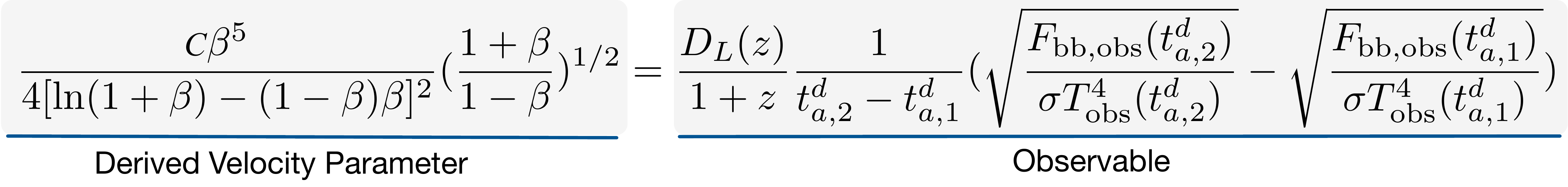}
\caption{Equation to compute the velocity from the thermal component: this equation is summarized from \citet{2018ApJ...852...53R}. The left-hand side term is only a function of velocity $\beta$, the right-hand side term is only of the observables. $D_L(z)$ is the luminosity distance for redshift $z$. From the observed thermal flux $F_\mathrm{bb,obs}$ and temperature $T_\mathrm{obs}$ at  arrival times of the detector $t_{a,1}^d$ and $t_{a,2}^d$, the velocity and the corresponding Lorentz factor can be computed. This equation assumes uncollimated emission and considers only the radiation coming from the line of sight. The computed velocity is instantaneous and there is no reliance on the expansion history.}
\label{fig:funcV}
\end{figure*}

{
There is finally a third additional extended thermal mission (ETE) observed in BdHNe and in the  the X-ray flares \citep{2018ApJ...852...53R}, this ETE has allowed the determination of the velocity of expansion and Lorentz Gamma factor of the thermal emission based on the variation in time of the observed radius and temperature of the thermal emission (see equation in Fig.~\ref{fig:funcV}) under the assumption of uncollimated emission and considering only the radiation coming from the line of sight. The left-hand side term is only a function of the velocity $\beta$, the right-hand side term is only function of the observables, $D_L(z)$ is the luminosity distance for redshift $z$. Therefore, from the observed thermal flux $F_\mathrm{bb,obs}$ and temperature $T_\mathrm{obs}$ at times $t_1$ and $t_2$, we can compute the velocity $\beta$. This highly non-linear equation is not straightforwardly solvable analytically so in the present paper we solve it numerically after verifying the monotonically increasing behavior of the left-hand side term as a function of $\beta$ (see, e.g., Bianco, Rueda, Ruffini, Wang, in preparation).
}

{
The second progress has been presented in \citet{2016ApJ...833..107B} and more recently in \citet{2018arXiv180304356B}: the first 3D SPH simulations of the IGC leading to a BdHN are there presented. We simulate the SN explosion of a CO$_{\rm core}$ forming a binary system with a NS companion. We follow the evolution of the SN ejecta, including their morphological structure, subjected to the gravitational field of both the new NS ($\nu$NS), formed at the center of the SN, and the one of the NS companion. We compute the accretion rate of the SN ejecta onto the NS companion as well as onto the $\nu$NS from SN matter fallback. We determine the fate of the binary system for a wide parameter space including different CO$_{\rm core}$ masses, orbital periods ($\sim 300$~s) and SN explosion geometry and energies. We evaluate, for selected NS equations of state, if the accretion process leads the NS either to the mass-shedding limit, or to the secular asymmetric instability for gravitational collapse to a BH, or to a more massive, fast rotating, but stable NS. We also assess whether the binary keeps or not gravitationally bound after the SN explosion, hence exploring the space of binary and SN explosion parameters leading to the formation of $\nu$NS-NS or $\nu$NS-BH binaries. The consequences of our results for the modeling of GRBs via the IGC scenario are discussed in \citet{2018arXiv180304356B}. The relevance of these simulations for GRB 151027A which is subject of this paper will be illustrated below, see Fig.~\ref{fig:cc}.
}

\begin{figure}
\centering
\includegraphics[width=1\hsize,clip]{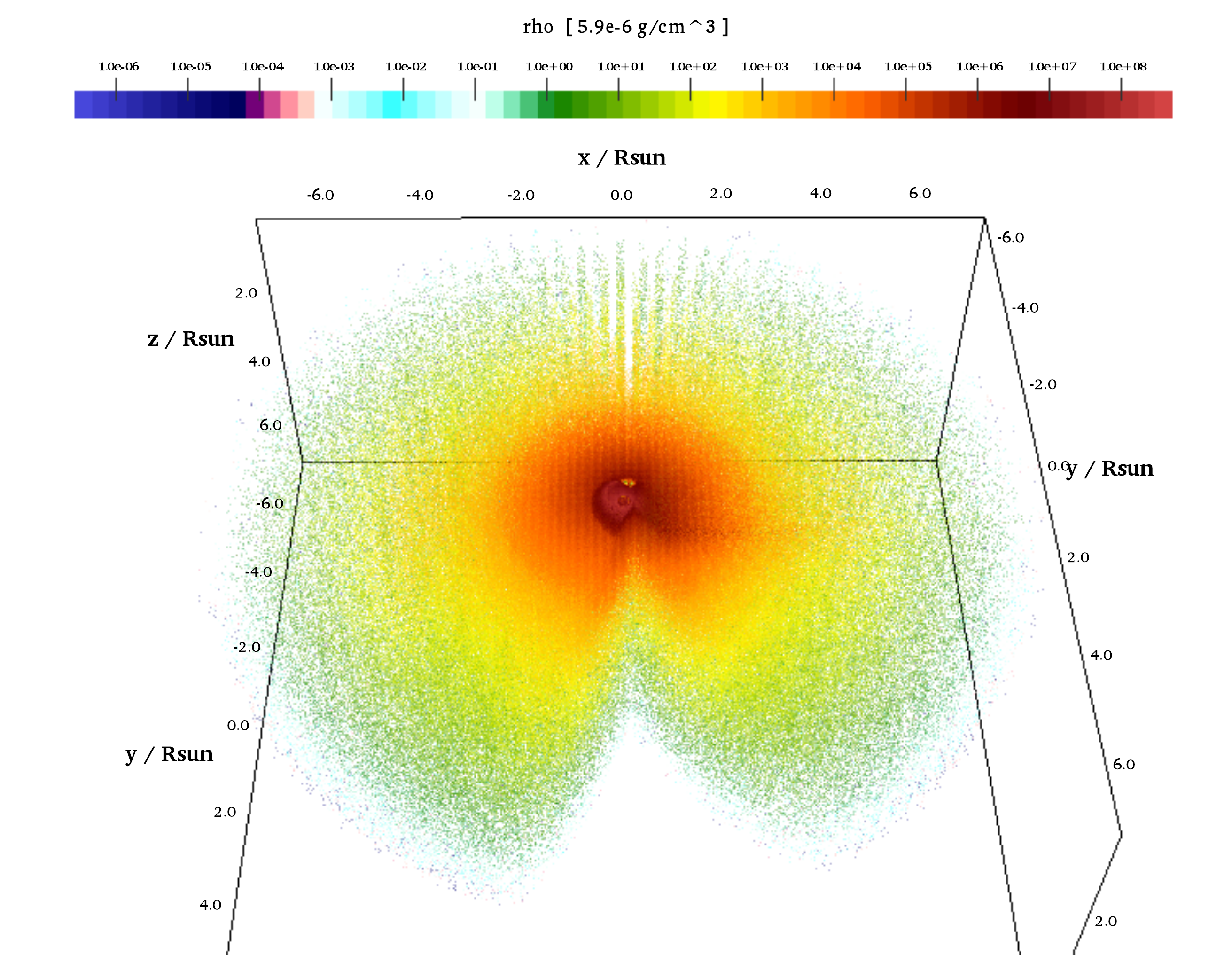}
\caption{Three-dimensional, half-hemisphere view of the density distribution of the SN ejecta at the moment of BH formation in a BdHN. The simulation is performed with an SPH code that follows the SN ejecta expansion under the influence of the gravitational field of both the $\nu$NS formed at the center of the SN and of the NS companion. It includes the effects of the orbital motion and the changes in the NS gravitational mass by the hypercritical accretion process \citep[see][for additional details]{2016ApJ...833..107B}. The binary parameters of this simulation are: the NS companion has an initial mass of $2.0~M_\odot$; the CO$_{\rm core}$, obtained from a progenitor with ZAMS mass $M_{\rm ZAMS}=30~M_\odot$, leads to a total ejecta mass $7.94~M_\odot$ and to a $1.5~M_\odot$ $\nu$NS, the orbital period is $P\approx 5$~min (binary separation $a\approx 1.5\times 10^{10}$~cm). {Only the sources, whose ultra-relativistic emission lies within the allowed cone of $\sim 10^\circ$ with low baryon contamination, will trigger the gamma-ray instrument (e.g. Fermi/GBM or Swift/BAT).}}
\label{fig:cc}
\end{figure}

\begin{figure}[ht]
\centering
\includegraphics[width=\hsize,clip]{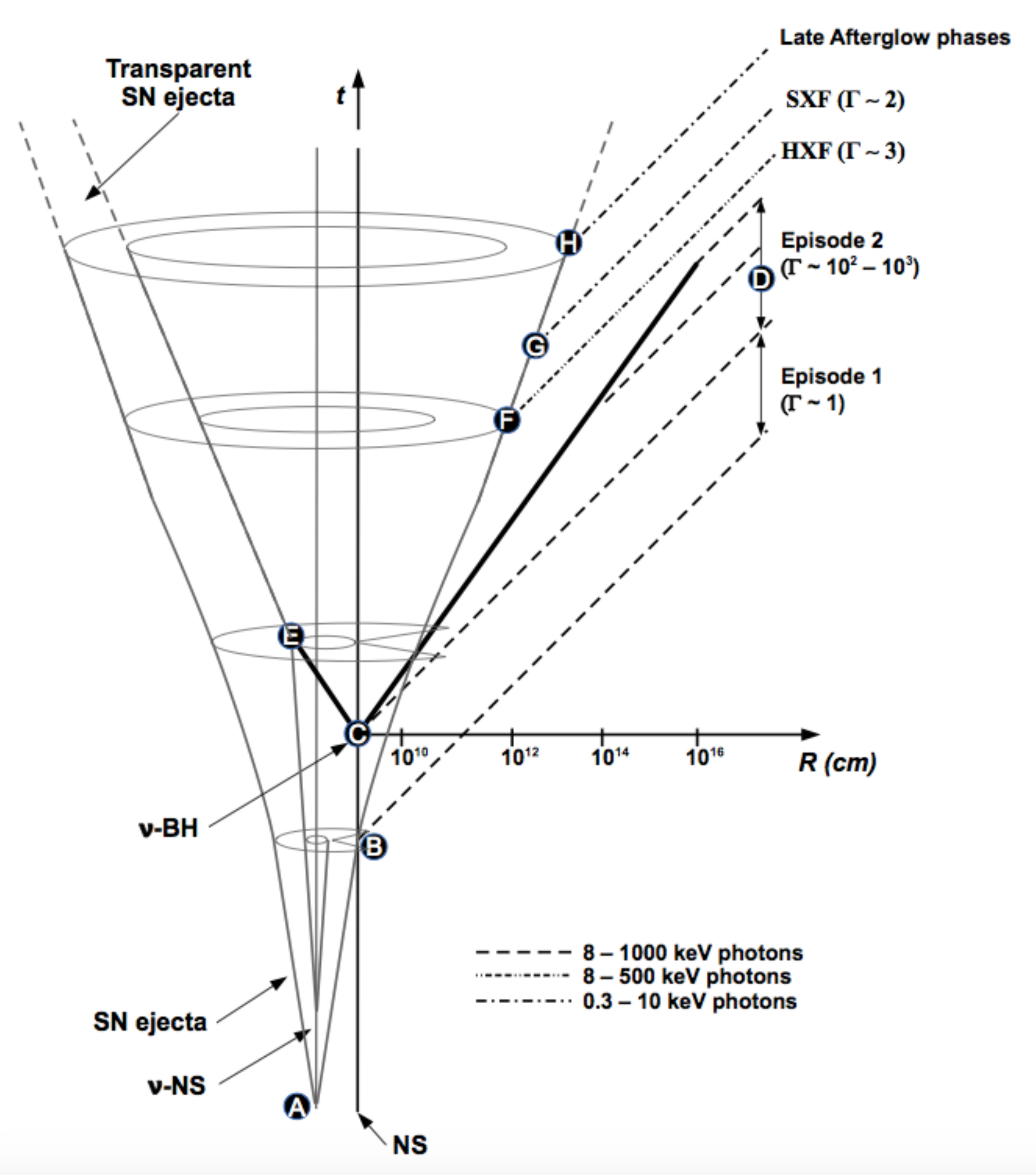}
\caption{Space-time diagram (not in scale) of BdHNe. The CO$_\mathrm{core}$ explodes as a SN at point A and forms a $\nu$NS. The companion NS (bottom right line) accretes the SN ejecta starting from point B, giving rise to the non-relativistic Episode 1 emission (with Lorentz factor $\Gamma\approx 1$). At the point C the NS companion collapses into a BH, and an $e^+e^-$ plasma --- the dyadosphere --- is formed \citep{RSWX2}. The following self-acceleration process occurs in a spherically symmetric manner (thick black lines). A large portion of plasma propagates in the direction of the line of sight, where the environment is cleaned up by the previous accretion into the NS companion, finding a baryon load $B \lesssim 10^{-2}$ and leading to the GRB UPE gamma-ray spikes (Episode 2, point D) with $\Gamma \sim 10^2$--$10^3$. The remaining part of the plasma impacts with the high density portion of the SN ejecta (point E), propagates inside the ejecta encountering a baryon load $B \sim 10^{1}-10^2$, and finally reaches transparency, leading to the hard X-ray flare emission (point F) in gamma rays with an effective Lorentz factor $\Gamma \lesssim 10$ and to soft X-ray flare emission (point G) with an effective $\Gamma \lesssim 4$, which are then followed by the late afterglow phases (point H). For simplicity, this diagram is 2D and static and does not attempt to show the 3D rotation of the ejecta.}
\label{fig:Carlo}
\end{figure}

{
Finally, we present an update of the BdHN space-time diagram (see Fig.~\ref{fig:Carlo}) which clearly evidences the large number of episodes and physical processes, each with observationally computed time-varying Lorentz $\Gamma$ factors, which require the systematic use of the four different time coordinates, already indicated in \citet{Ruffini2001c}. The diagram illustrates departures from the traditional collapsar-fireball description of a GRB. The diagram shows how the sequence of events of the UPE, of the hard X-ray flare and of the soft X-ray flare occur in a sequence only when parametrized in the arrival time and are not in fact causally related.
}

{
We recall that within our model the line of sight of the protypical GRB 151027A lies in the equatorial plane of the progenitor binary system. The more general case of an arbitrary viewing angle has been explored in \citet{2018arXiv180305476R}, and some specific additional characteristic features common to the collapsar model have been manifested in this more general case.
}

\section{Ultra-relativistic Prompt Emission (UPE)}\label{sec2}

\begin{figure}[ht]
\centering
(a)\includegraphics[width=0.95\hsize,clip]{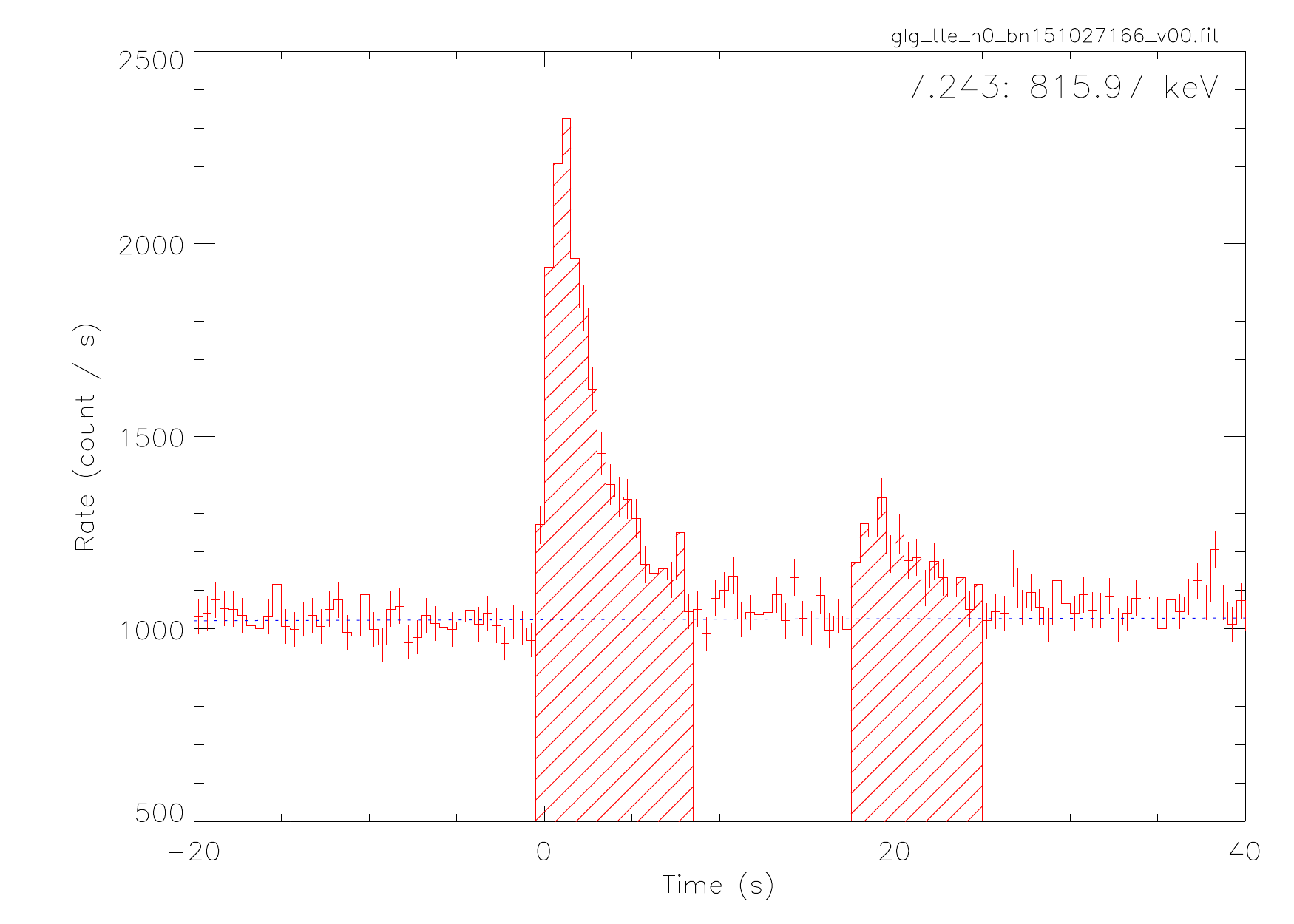}
(b)\includegraphics[width=0.95\hsize,clip]{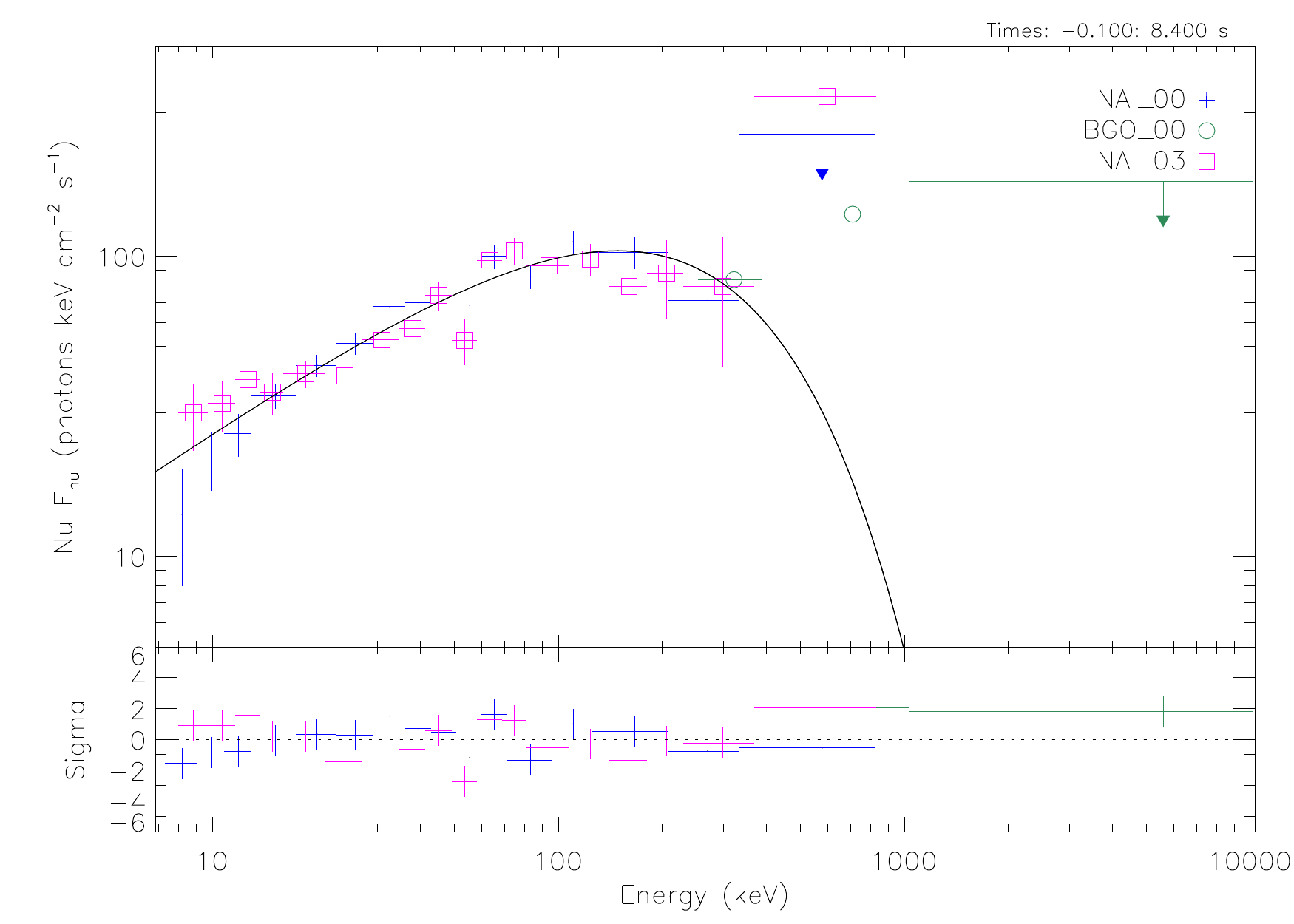}
(c)\includegraphics[width=0.95\hsize,clip]{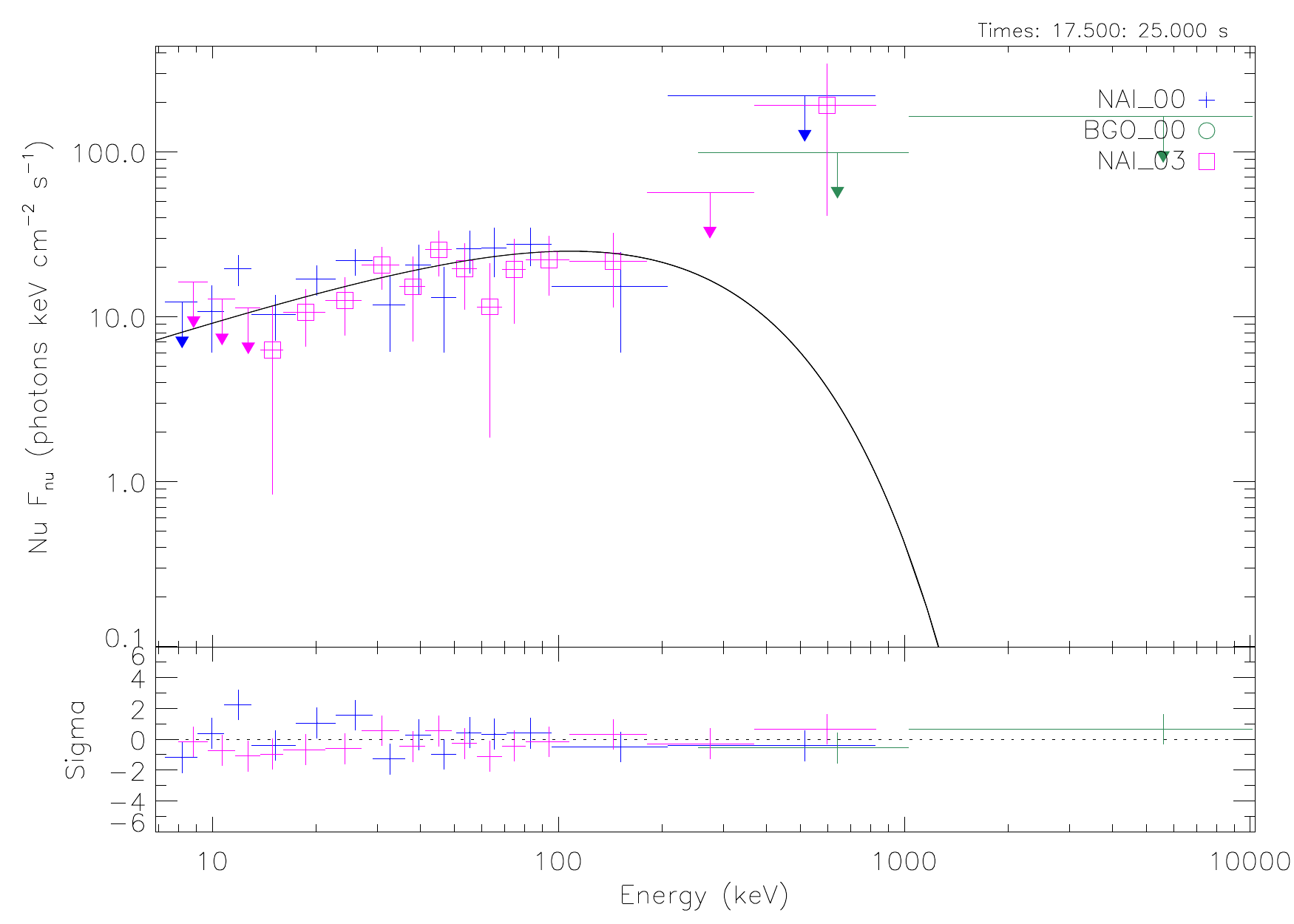}
\caption{(a) The \textit{Fermi}-GBM light curve from the NaI-n0 detector ($\approx8$--$800$~keV) of the UPE of GRB 151027A. The dotted horizontal line corresponds to the $\gamma$-ray background. (b) Time-integrated $\nu F_\nu$ spectrum of the first spike. (c) Time-integrated $\nu F_\nu$ spectrum of the second spike.}
\label{fig1a}
\end{figure}

\begin{figure*}
\centering
(a)\includegraphics[width=0.45\hsize,clip]{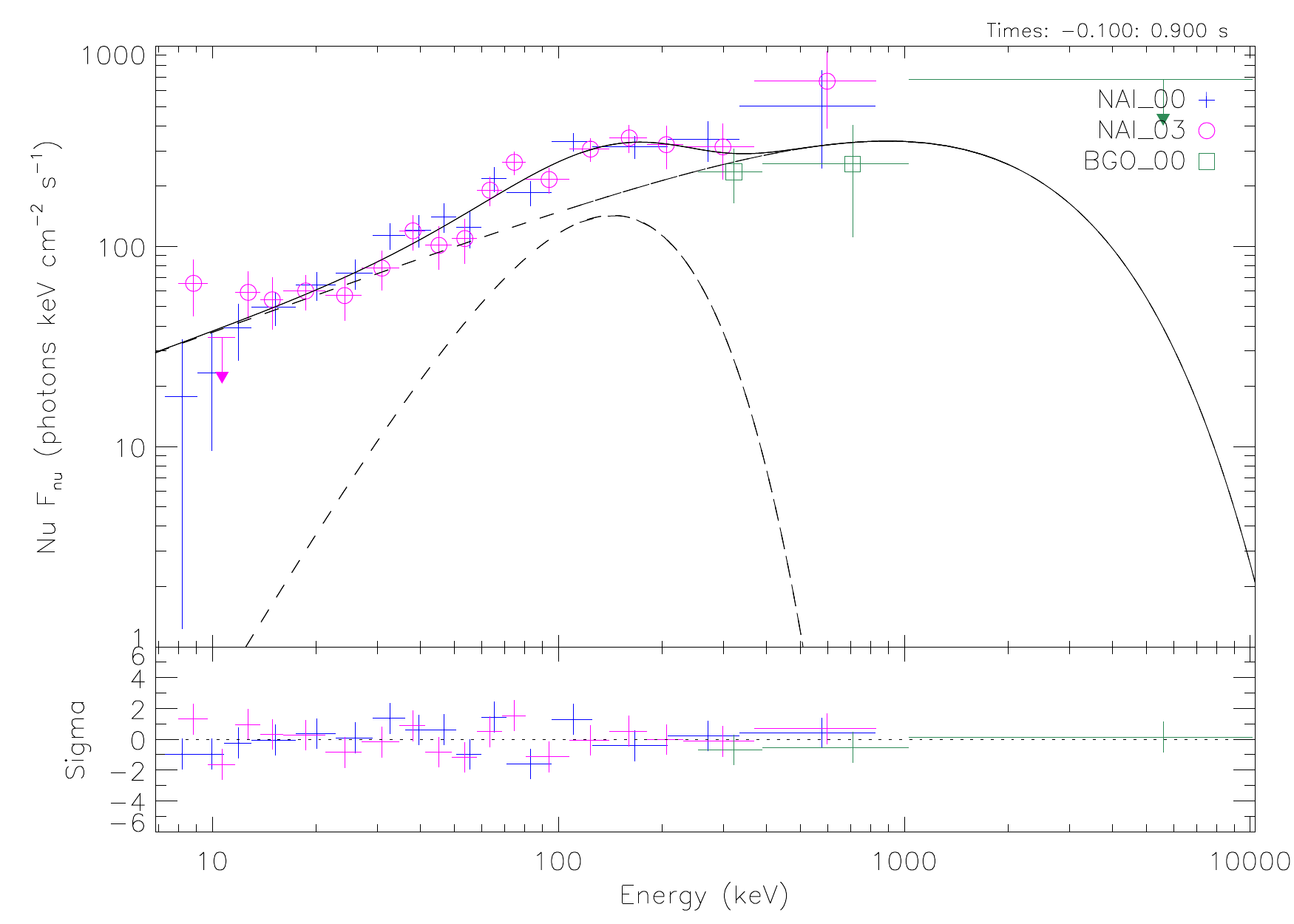}
(b)\includegraphics[width=0.45\hsize,clip]{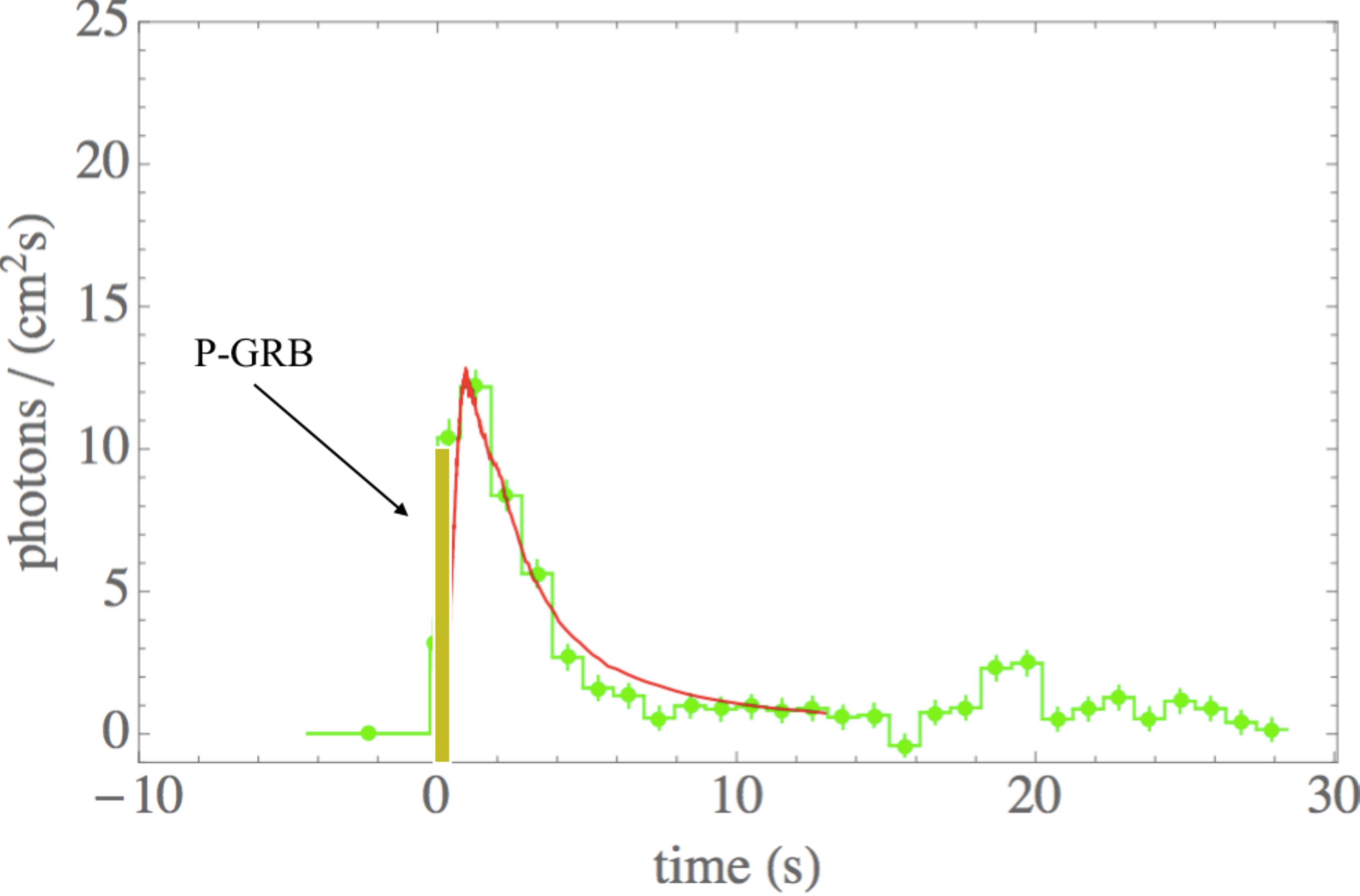}
(c)\includegraphics[width=0.45\hsize,clip]{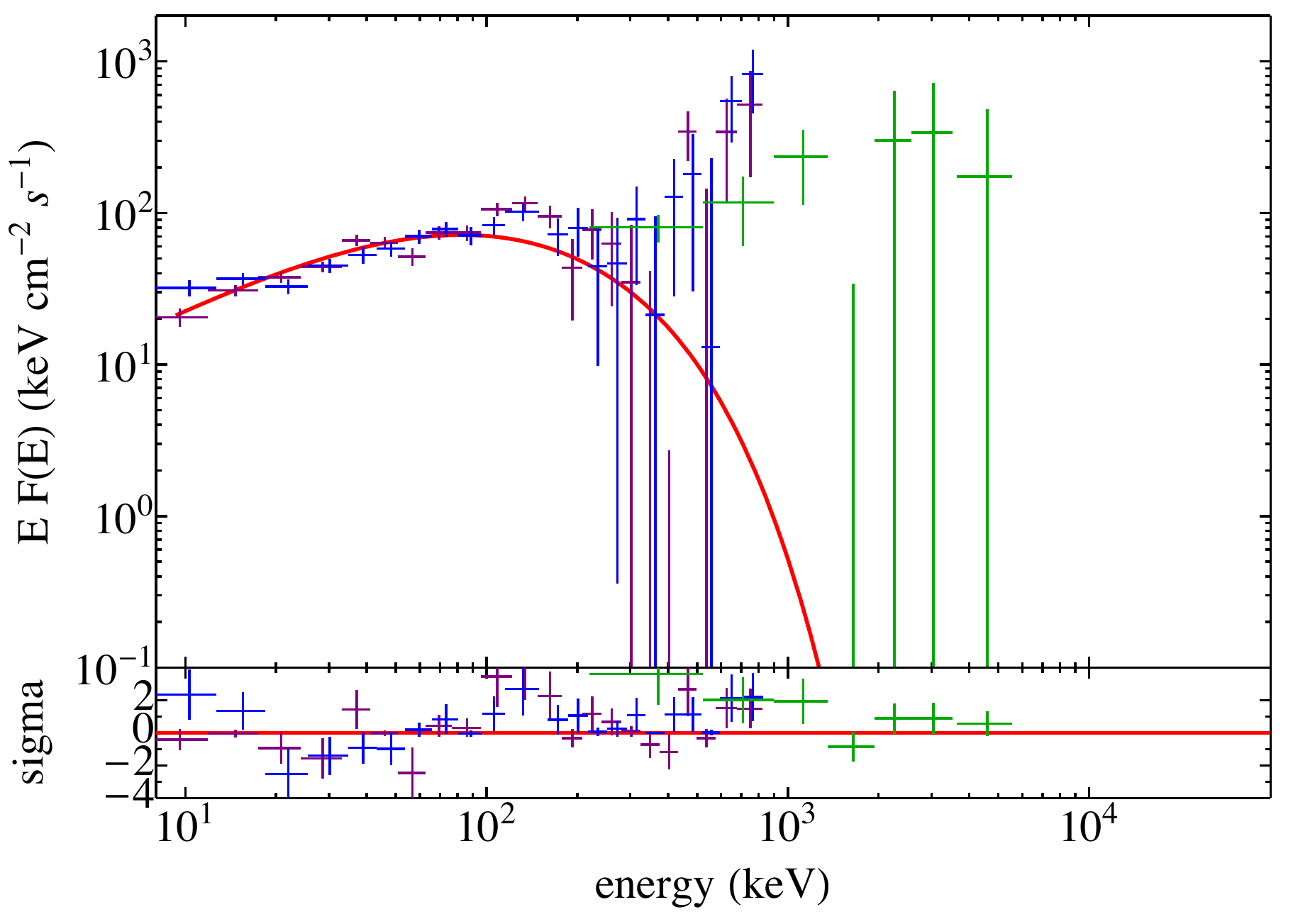}
(d)\includegraphics[width=0.45\hsize,clip]{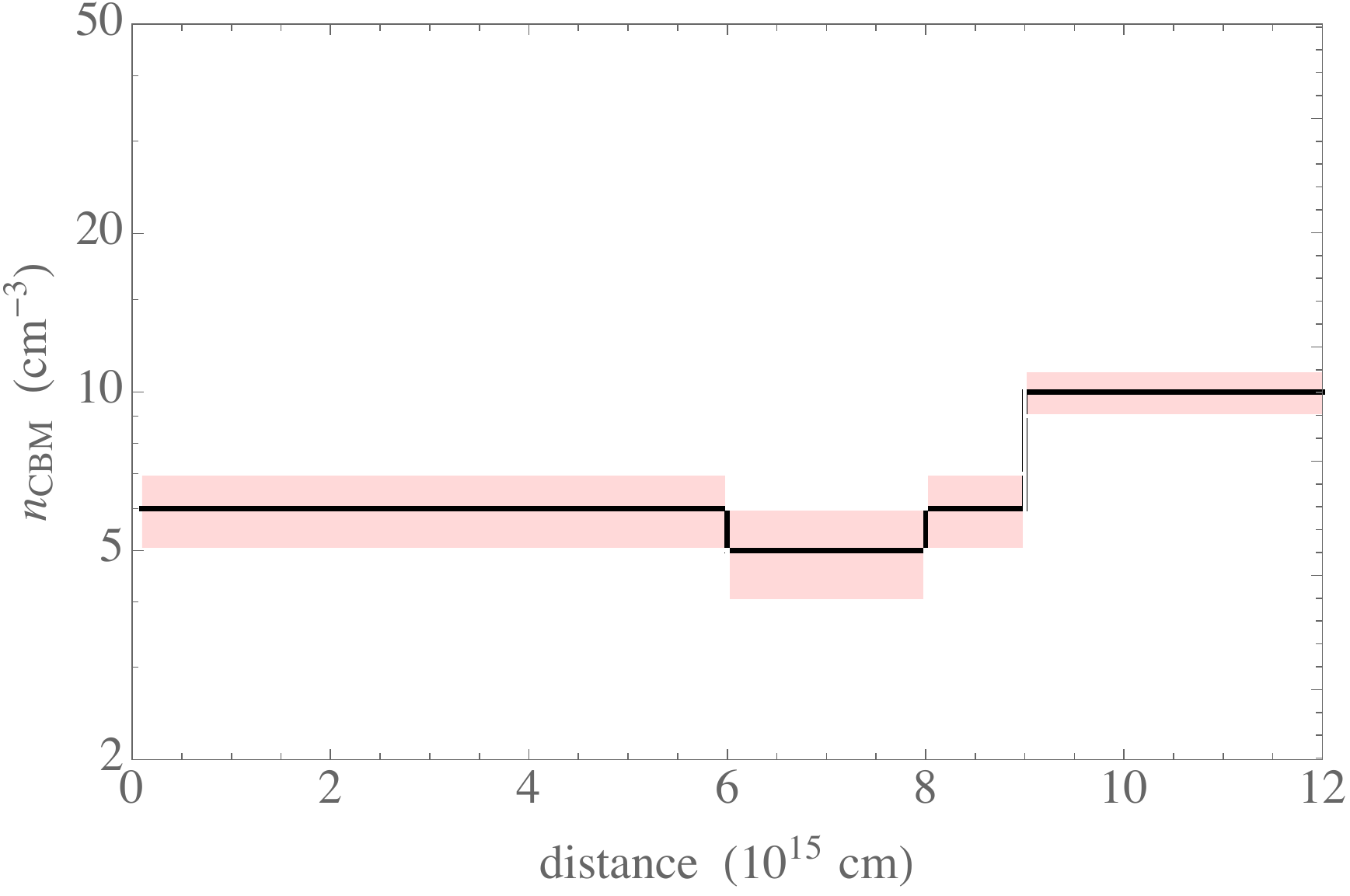}
\caption{Ultra-relativistic prompt emission (UPE): (a) The combined NaI-n0, n3+BGO-b0 $\nu F_\nu$ spectrum of the P-GRB in the time interval $T_0-0.1$--$T_0+0.9$~s. The best-fit model is CPL+BB. (b) The comparison between the background subtracted $10$--$1000$~keV \textit{Fermi}-GBM light curve (green) and the simulation with the fireshell model (red curve) in the time interval $T_0+0.9$--$T_0+9.6$~s. (c) The comparison between the NaI-n0 (purple squares), n3 (blue diamonds) and the BGO-b0 (green circles) $\nu F_\nu$ data in the time interval $T_0+0.9$--$T_0+9.6$~s and the simulated fireshell spectrum (red curve). (d) The radial density of the CBM clouds used for the above UPE light curve and spectrum simulations.}
\label{fig0a}
\end{figure*}

\begin{figure}
\centering
\includegraphics[width=1.1\hsize,clip]{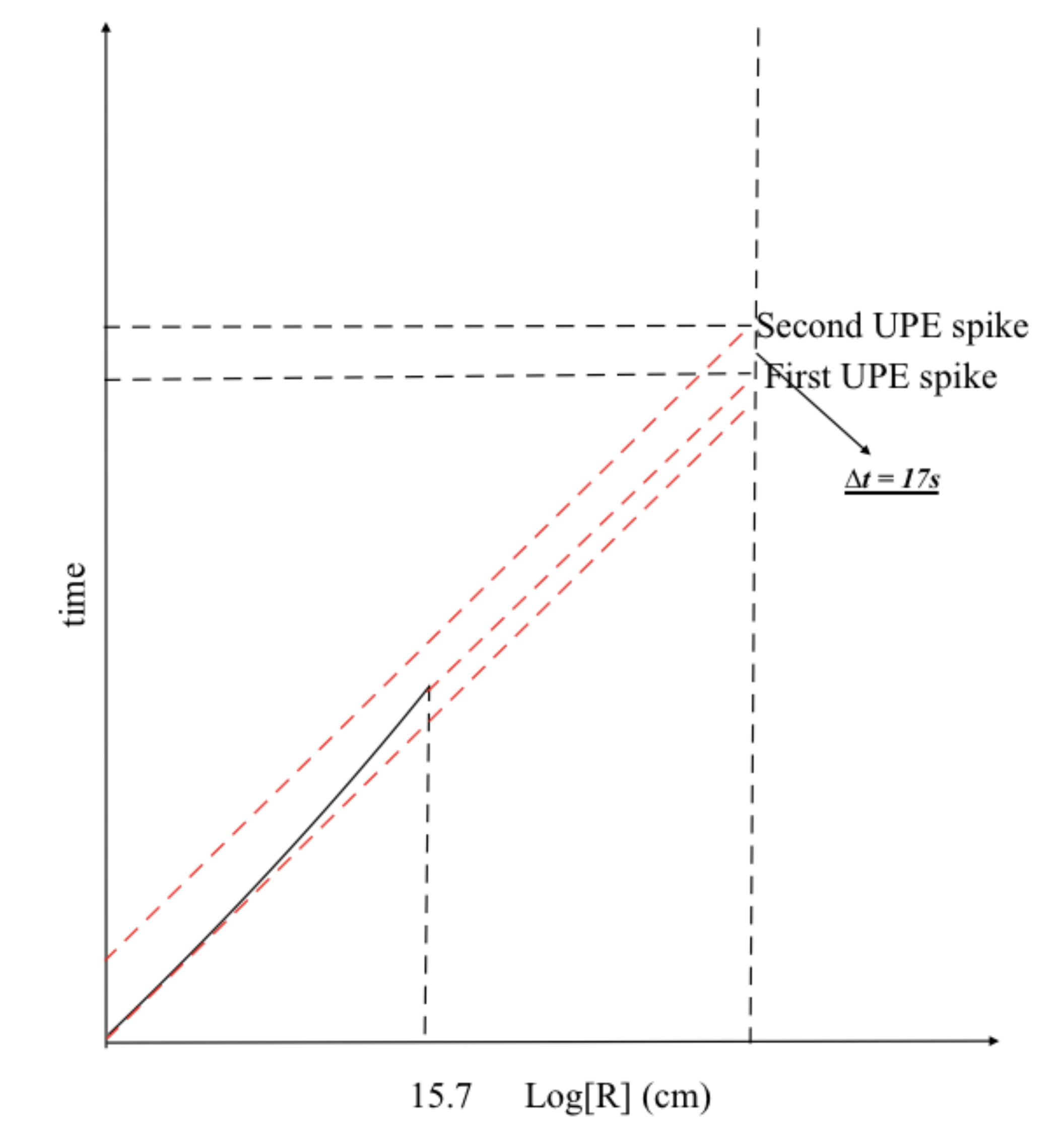}
\caption{{Spacetime diagram of the UPE. The initial $e^+e^-$ plasma self-accelerates in the small-density cone until it reaches transparency (curved black line), producing the first of the two ultra-relativistic UPE spikes (lower solid red line). The second one is produced by a latter emission from the BH formation, with a difference in the observed time of $\sim 17$~s (rest-frame $\sim 9.4~s$) (upper solid red line).}}
\label{fig:03}
\end{figure}

GRB 151027A was detected and located by the \textit{Swift} Burst Alert Telescope (BAT) \citep{2015GCN..18478...1M}. It was also detected by the \textit{Fermi} Gamma-ray Burst Monitor (GBM) \citep{2015GCN..18492...1T}, MAXI \citep{2015GCN.18525....1M} and by \textit{Konus}-Wind \citep{2015GCN..18516...1G}. The \textit{Swift} X-Ray Telescope (XRT) started its observation $87$~s after the burst trigger \citep{2015GCN..18482...1G}. The redshift of the source, measured through the MgII doublet in absorption from the Keck/HIRES spectrum, is $z=0.81$ \citep{2015GCN..18487...1P}. The LAT boresight of the source was $10^\mathrm{o}$ at the time of the trigger, there are no associated high energy photons; an upper limit of observed count flux is computed as $9.24 \times 10^{-6}$~photons~cm$^{-2}$~s$^{-1}$ following the standard Fermi-LAT likelihood analysis. The BAT light curve shows a complex peaked structure lasting at least $83$ seconds. XRT began observing the field $48$~s after the BAT trigger. The GBM light curve consists of various pulses with a duration of about $68$~s in the $50$--$300$~keV band. The Konus-Wind light curve consists of various pulses with a total duration of $\sim 66$~s. The MAXI detection is not significant, but the flux is consistent with the interpolation from the \textit{Swift}/XRT light curve. The first $25$~s (rest-frame $14$~s) corresponds to the UPE. It encompasses two spikes of duration $\approx8.5$~s and $\approx7.5$~s, respectively with a separation between two peaks $\approx17$~s (see Fig.~\ref{fig1a}~(a)). The rest-frame $1$--$10^4$~keV isotropic equivalent energies computed from the {time integrated} spectra of these two spikes (see Figs.~\ref{fig1a}~(b) and (c)) are $E_{\rm iso,1}=(7.26\pm0.36)\times10^{51}$~erg and $E_{\rm iso,2}=(4.99\pm0.60)\times10^{51}$~erg, respectively.

A similar analysis was performed by \cite{2017A&A...598A..23N}. They describe the two spikes of the UPE by a single light curve with a ``Fast Rise and Exponential Decay'' (FRED) shape.

{We analyze} the first spike (see Fig.~\ref{fig0a}) as the traditional UPE of a long GRB within the fireshell model \citep[see, e.g.,][for a review]{RVX}.

Thanks to the wide energy range of the \textit{Fermi}-GBM instrument ($8$--$1000$~keV) it has been possible to perform a time-resolved analysis within the UPE phase to search for the typical P-GRB emission at the transparency of the $e^+e^-$--baryon plasma \citep{RSWX2,2000A&A...359..855R,Ruffini2001}. Indeed, we find this thermal spectral feature in the time interval $T_0-0.1$--$T_0+0.9$~s (with respect to the \textit{Fermi}-GBM trigger time $T_0$). The best-fit model of this emission is a composition of a black-body (BB) spectrum and a cut-off power-law model (CPL, see Fig.~\ref{fig0a}(a)). The BB component has an observed temperature $kT=(36.6\pm5.2)$~keV and an energy $E_{\rm BB}=(0.074\pm0.038)\times E_{\rm iso,1}=(5.3\pm2.7)\times10^{50}$~erg. These values are in agreement with an initial $e^+e^-$ plasma of energy $E_{\rm iso,1}$, with a baryon load $B=(1.92\pm0.35)\times10^{-3}$, and a Lorentz factor and a radius at the transparency condition of $\Gamma_0=503\pm76$ and $r_{\rm tr}=(1.92\pm0.17)\times10^{13}$~cm, respectively.

{We turn now to the simulation of the remaining part of the first spike of the UPE} (from $T_0+0.9$~s to $T_0+9.6$~s). In the fireshell model, this emission occurs after the P-GRB and results from {the slowing down of the accelerated baryons due to their interaction with the CBM} \citep{2002ApJ...581L..19R,Ruffini2006,Patricelli2011}. To simulate the UPE light curve and its corresponding spectrum, we need to derive the number density of the CBM clouds surrounding the burst site. The agreement between the observations and the simulated light curve (see Fig.~\ref{fig0a}(b)) and the corresponding spectrum (see Fig.~\ref{fig0a}(c)) is obtained for an average CBM density of $(7.46\pm1.2)$~cm$^{-3}$ (see Fig.~\ref{fig0a}(d)) consistent with the typical value of the long burst host galaxies at radii $\simeq 10^{16}$~cm. By contrast the second spike of the UPE appears to be featureless. 

{
The general conclusion of the UPE is the following:}
{
From the morphological 3D simulation, the SN ejecta is distorted by the binary accretion: a cone of very low baryon contamination is formed along the direction from the SN center pointing to the newly born BH, see Fig.~\ref{fig:cc}. A portion of $e^+e^-$ plasma generated from the BH formation propagates through this cone and engulfs a low baryon load of $B=(1.92\pm0.35)\times10^{-3}$ and reaching a Lorentz gamma factor of $\Gamma_0=503\pm76$. The $e^+e^-$ plasma  self-accelerates and expands ultra-relativistically till reaching transparency \citep{1998bhhe.conf..167R,2007PhRvL..99l5003A,2010PhR...487....1R}, when a short duration ($<1$~s) thermal emission occurs: the P-GRB. The ultra-relativistic associated baryons then interact with the circumburst medium (CBM) clouds: the dynamics of the plasma has been integrated by the classical hydrodynamics equations, by the equation of annihilation-creation rate \citep{2001A&A...368..377B,Bianco2004,Bianco2005a,2005ApJ...633L..13B,2006ApJ...644L.105B}, and it enables to simulate the structure of spikes in the prompt emission, and it has been applied to the case of BdHNe \citep[see, e.g.,][]{2002ApJ...581L..19R,2005ApJ...634L..29B,2012A&A...543A..10I,2012A&A...538A..58P,2013A&A...551A.133P,2016ApJ...831..178R}. For typical baryon load for the cone direction, $10^{-4} \lesssim B \lesssim 10^{-2}$, leading to a Lorentz factor $\Gamma \approx 10^2$--$10^3$, characteristic the prompt emission occurs in a distance  $\approx10^{15}$--$10^{17}$~cm from the BH \citep{2016ApJ...832..136R}.}
{
\begin{enumerate}
\item 
a  double emission is  clearly manifested by presence of  the two spikes at the time interval of the $17 s$ (rest-frame $9$ s). We are currently examining the possibility that this double emission is an imprinting of the process of the BH formation.
\item 
when we take into account the rotation period of the binary $\sim 300$~s we see that UPE occurs in a cone centered in the BH of $10^{\circ}$; 
\item 
this conical region is endowed with very low density determined by the P-GRB and the inferred CBM medium density of $(7.46\pm1.2)$~cm$^{-3}$ up to $10^{16}$ cm from BH  along the cone, see Fig.~\ref{fig0a}(d)
\end{enumerate}
}
{
This conceptual framework can in principle explain the featureless nature of the second spike which propagates along the region which has already been swept by the first spike (see Fig.~\ref{fig:03}).}

\section{Hard and Soft X-ray flare }\label{sec3}

\subsection{Hard X-ray flare}

\begin{figure}[ht]
\centering
(a)\includegraphics[width=\hsize,clip]{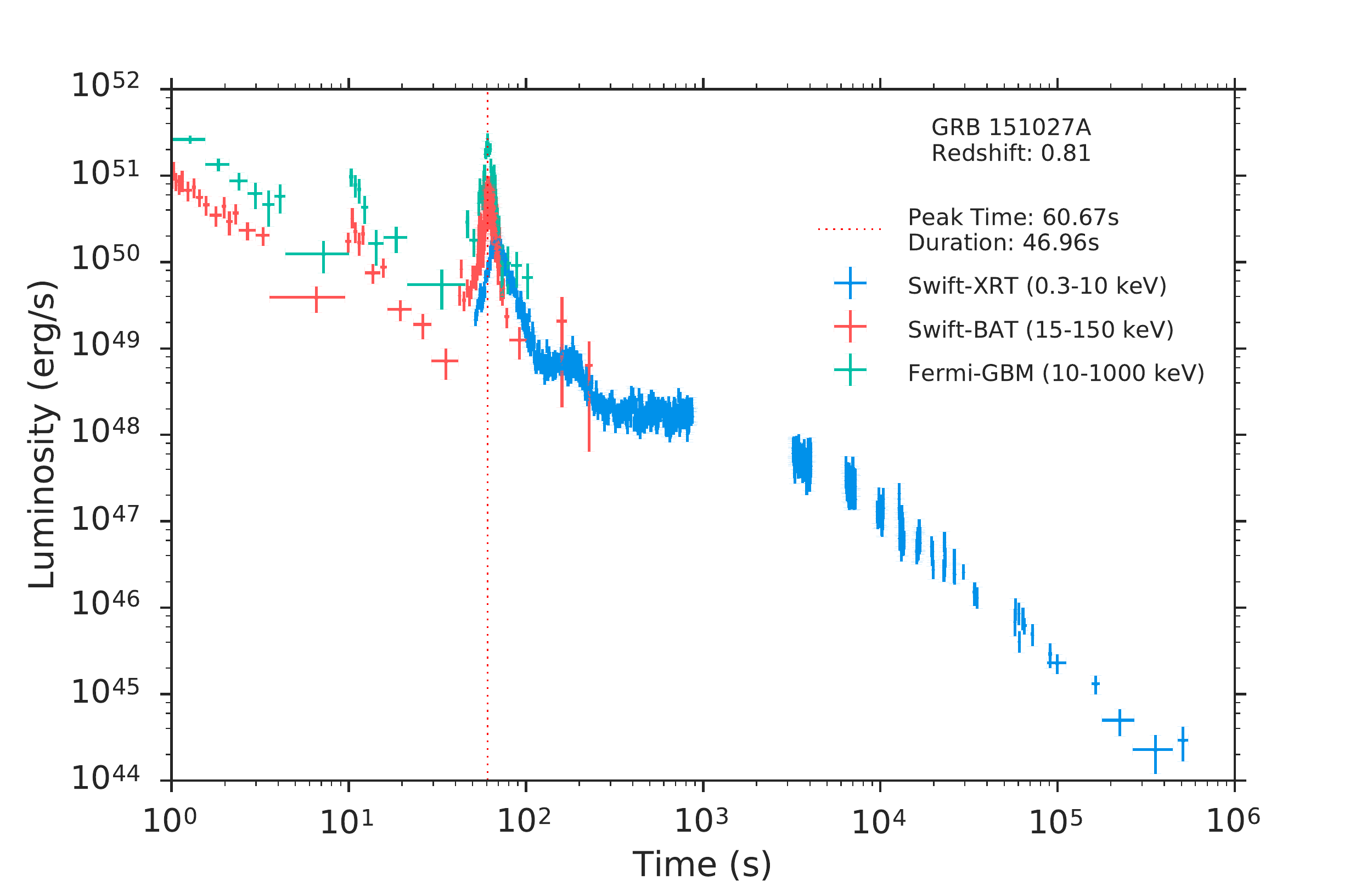}
(b)\includegraphics[width=0.88\hsize,clip]{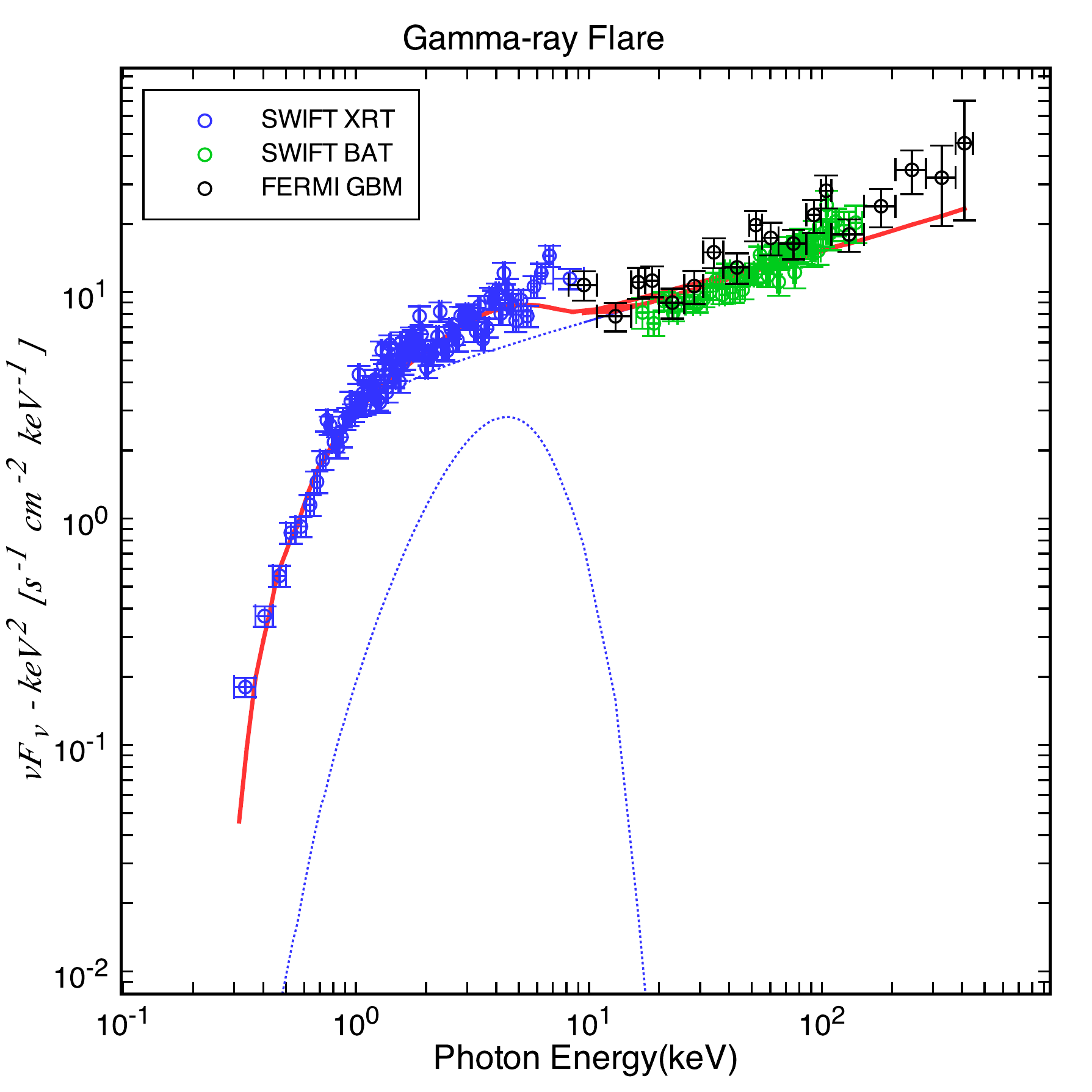}
\caption{(a) Luminosity light curves in the rest-frame energy bands $10$--$1000$~keV for \textit{Fermi}-GBM (green), $15$--$150$~keV for \textit{Swift}-BAT (red), and $0.3$--$10$~keV for \textit{Swift}-XRT (blue). The red dotted line marks the position of the hard X-ray flare. (b) Time-integrated $\nu F_\nu$ spectrum of the hard X-ray flare and the PL+BB model (solid red curve) best-fitting the data.}
\label{fig1b}
\end{figure}

{We turn now to the hard X-ray flare and the soft X-ray flare.} The hard X-ray flare is observed in the time interval $94$--$180$~s (corresponding to the rest-frame time interval $52$--$99$~s, see Fig.~\ref{fig1b}~(a)). The luminosity light curves in the rest-frame energy bands $10$--$1000$~keV for \textit{Fermi}-GBM (green), $15$--$150$~keV for \textit{Swift}-BAT (red), and $0.3$--$10$~keV for \textit{Swift}-XRT (blue) are displayed. The total isotropic energy of the hard X-ray flare is $E_{\gamma}=(3.28\pm0.13)\times10^{52}$~erg. The overall spectrum is best-fit by a superposition of a power-law (PL) function with an index $-1.69\pm0.01$ and a BB model with a temperature $kT=1.13\pm0.08$~keV (see Fig.~\ref{fig1b}~(b)).

\begin{figure*}
\centering
\includegraphics[width=\hsize,clip]{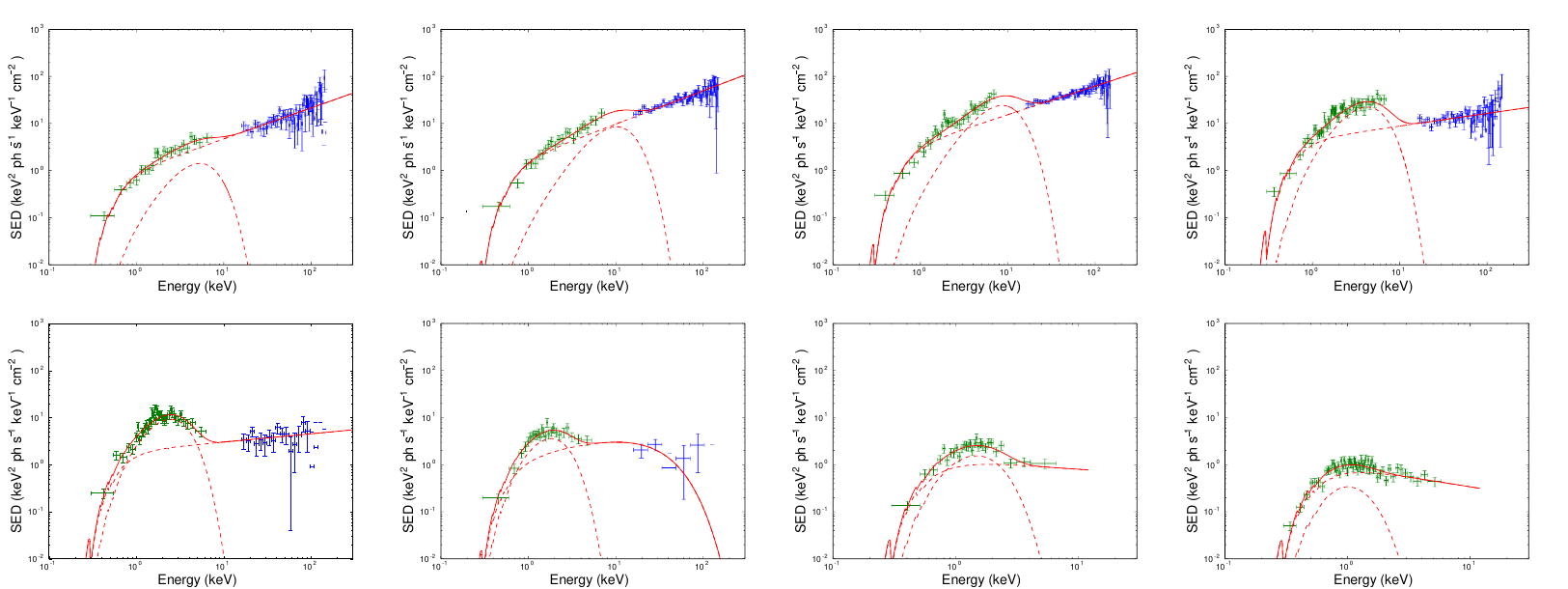}
\caption{hard X-ray flare: Time-resolved $\nu$F$_\nu$ spectra of the $8$ time intervals in Tab.~\ref{tab1} (from the top left to the right and from the bottom left to the right). XRT data are displayed in green and BAT data in blue; BAT data points with no vertical lines corresponds to upper limits. Plots correspond to parameters obtained from minimum $\chi^2$ fit.}
\label{fig4}
\end{figure*}

We perform a more detailed analysis by dividing the whole hard X-ray flare duration ($94$--$180$~s) into $8$ intervals (indicated with $\Delta t_a^d$ in Tab.~\ref{tab1}). Among these time intervals, the first $6$ have both BAT and XRT data (total energy range $0.3$--$150$~keV), while the last $2$ fits involve XRT data only (energy range $0.3$--$10$~keV). The XRT data were extremely piled-up and corrections have been performed in a conservative way to ascertain that the BB is not due to pile-up effects \citep{2006A&A...456..917R}. The absorption of the spectrum below $2$~keV has been also taken into due account. We use here the following spectral energy distributions to fit the data: power-law (PL), CPL, PL$+$BB and CPL$+$BB. An extra BB component is always preferred to the simple PL models and, only in the sixth interval, to the CPL model whose cutoff energy may be constrained within $90$\% significance. The results of the time-resolved analysis are shown in Fig.~\ref{fig4} and summarized in Tab.~\ref{tab1}. The BB parameters and errors in Tab.~\ref{tab1} correspond, respectively, to the main values and the $90$\% probability interval errors with respect to the central values, both obtained from Markov Chain-Monte Carlo method applied in \texttt{XSpec} with $10^5$ steps (excluding first $10^4$). The values are in line with the ones corresponding to minimum $\chi^2$ and errors to the ones corresponding to intervals obtained from the difference $\Delta\chi^2=2.706$ from the minimum $\chi^2$ value. The only exception is the first time bin where $\chi^2_{min}$ value is almost two times lower than the main value. It is useful to infer the bulk Lorentz factor of the hard X-ray flare emission from the non-thermal component of the spectrum. Using the \textit{Fermi} data, the best-fit model for this non-thermal component in the time interval $95$--$130$~s is a CPL with a spectral cutoff energy $E_c=926\pm238$~keV. Such a cutoff can be caused by $\gamma\gamma$ absorption, for which the target photon's energy is comparable to $E_c$, i.e., $E_c\gtrsim[\Gamma m_e c^2/(1+z)]^2/E_c$ and, therefore, the Lorentz factor can be deduced by
\begin{equation}
\Gamma\approx\frac{E_c}{m_e c^2}(1+z)\,,
\label{gammamax}
\end{equation}
where $m_e$ is the electron mass. From the above value of $E_c$, we infer $\Gamma=3.28\pm0.84$, which represents an average over the hard X-ray flare duration. It is in the range of the ones observed in thermal component (see the first five columns of the Tab.~\ref{tab1}), coinciding in turn with the numerical simulation of the interaction of the $e^+e^-$ plasma with the SN ejecta described in the Sec.~\ref{sec5}.

\subsection{Soft X-ray flare}

\begin{figure}[ht]
\centering
(a)\includegraphics[width=\hsize,clip]{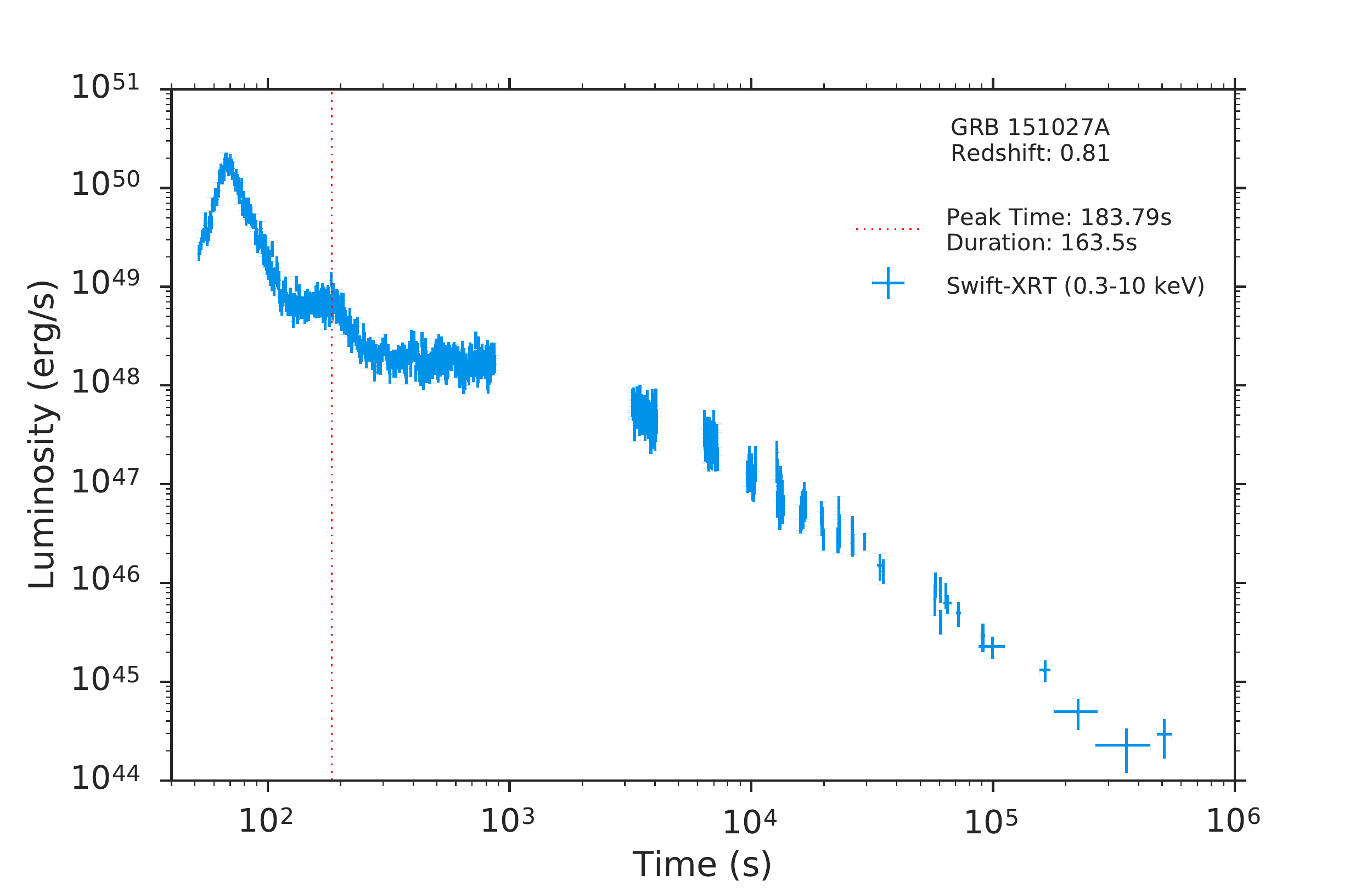}
(b)\includegraphics[width=0.88\hsize,clip]{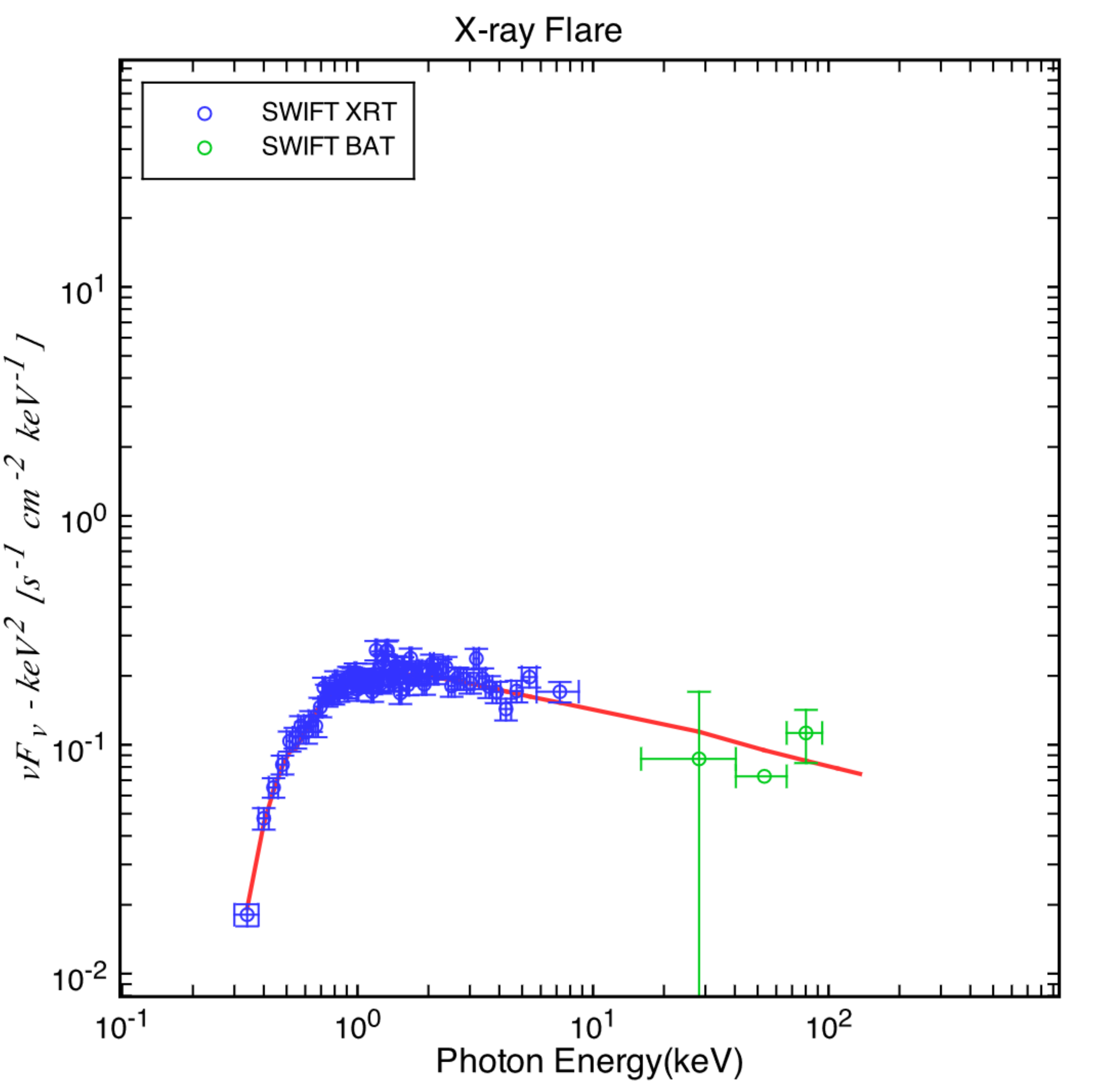}
\caption{(a)Rest-frame $0.3$--$10$~keV luminosity light curve of GRB 151027A. The red dotted line marks the position of the soft X-ray flare.  (b) Time-integrated $\nu F_\nu$ spectrum of the X-ray flare and the PL model (solid red curve) best-fitting the data.}
\label{fig1c}
\end{figure}

The soft X-ray flare, which has been discussed in \citet{2018ApJ...852...53R}, peaks at a rest-frame time $t_p=(184\pm 16)$~s, has a duration $\Delta t=(164\pm30)$~s, a peak luminosity $L_p=(7.1\pm 1.8)\times 10^{48}$~erg/s, and a total energy in the rest-frame $0.3$--$10$~keV energy range $E_X=(4.4 \pm 2.9)\times 10^{51}$~erg. The overall spectrum within its duration $\Delta t$ is best-fit by a PL model with a power-law index of $-2.24\pm0.03$ (see Fig.~\ref{fig1c}).

\begin{figure*}
\centering
\includegraphics[width=1\hsize,clip]{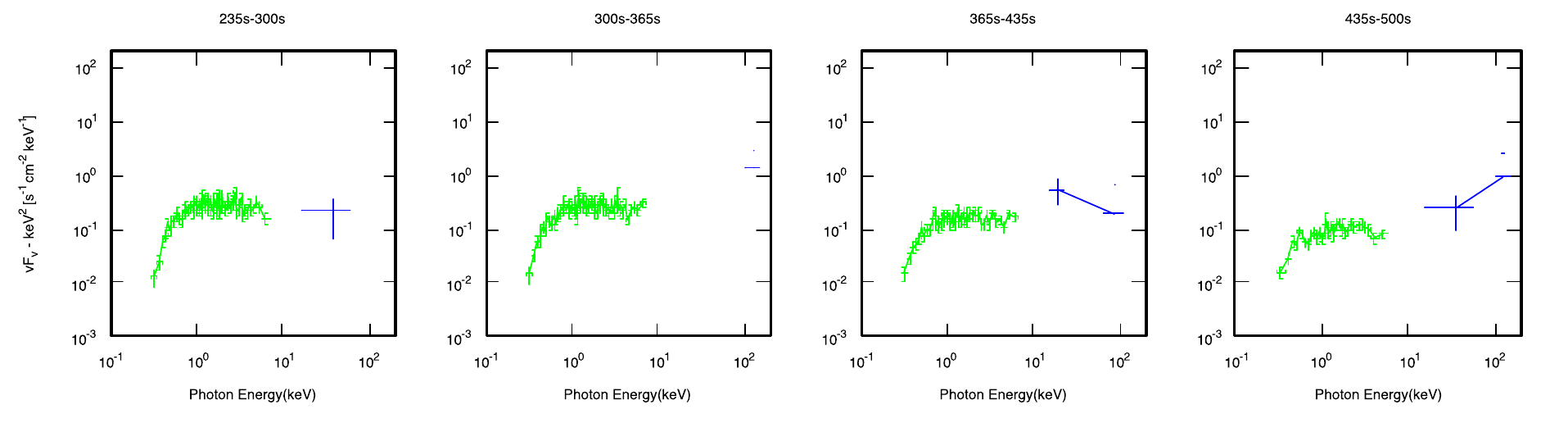}
\caption{soft X-ray flare: Time-resolved BAT (blue) and XRT (green) $\nu$F$_\nu$ spectra of the soft X-ray flare in the indicated time intervals.}
\label{figx}
\end{figure*}

We perform here also a time-resolved analysis of the soft X-ray flare. We divide the total interval $\Delta t$ into four sub-intervals, i.e., $235$--$300$~s, $300$--$365$~s, $365$--$435$~s and $435$--$500$~s in the observer frame (see Fig.~\ref{figx}). The best-fits of each of these $4$ time intervals are PL models with indexes ranging from $-2.3$ to $-2.1$, which are consistent with the typical values inferred in \citep{2018ApJ...852...53R}.

\begin{figure}
\centering
\includegraphics[width=1.1\hsize,clip]{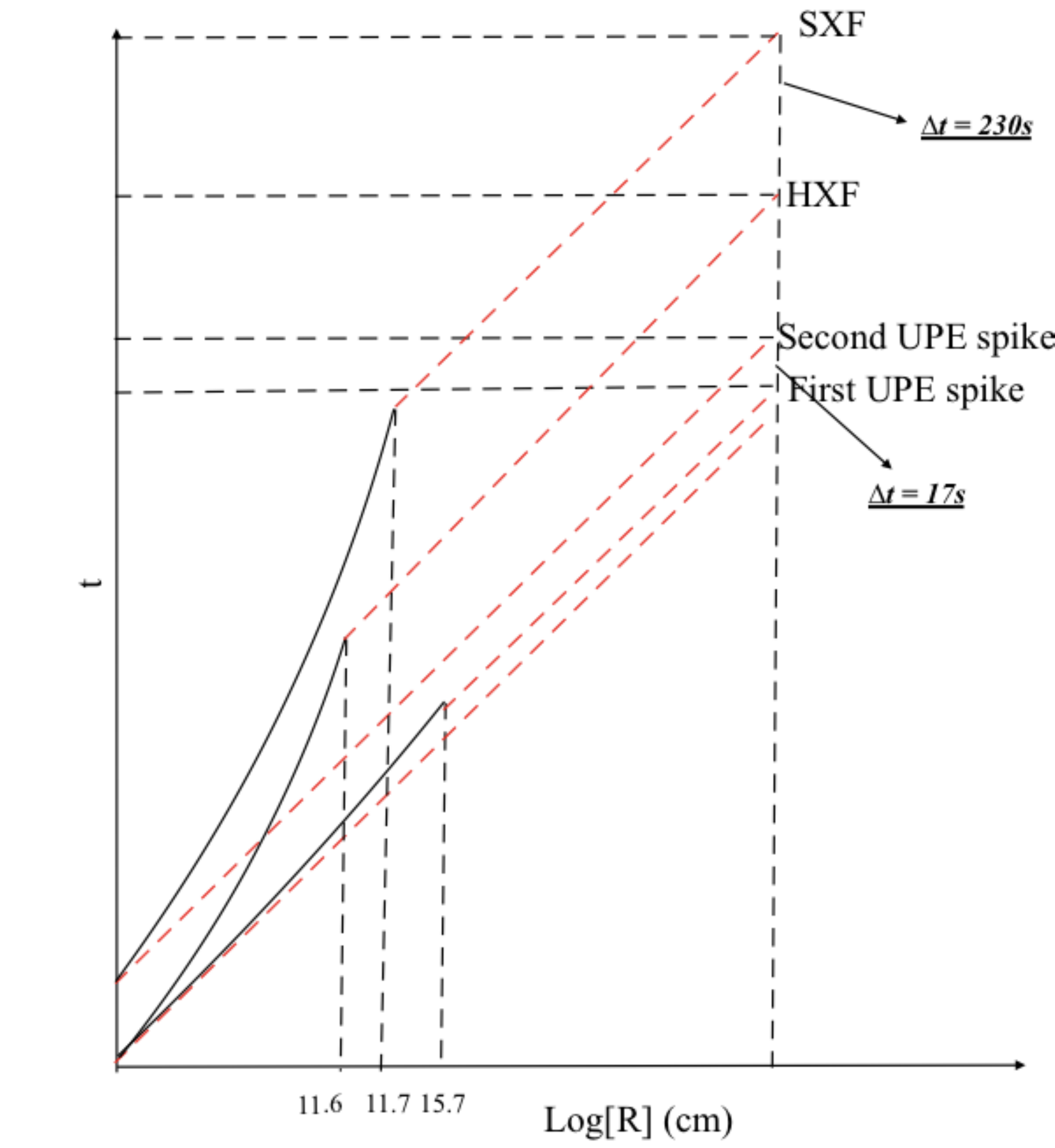}
\caption{{Same as Fig.~\ref{fig:03}, this time showing also the position of the plasma shock within the SN ejecta (dashed black lines) for each of the components of the UPE, until breakout. The first spike originates the hard X-ray flare and the second spike originates the soft X-ray flare. The photon wordlines (solid red lines) of hard X-ray flare and soft X-ray flare are observed with a time difference of $\sim 230$~s (rest-frame $\sim 130$~s ) due to the differential deceleration of the two UPE components within the SN ejecta.}}
\label{fig:03b}
\end{figure}

The complete space-time diagram, showing UPE, hard X-ray flare and soft X-ray flare, is represented in Fig.~\ref{fig:03b}.

\section{Evolution of thermal component around the hard X-ray flare}\label{sec4}

\begin{table*}
\centering
\caption{Hard X-ray flare: parameters of the time-resolved spectral analysis. Columns list, respectively, the time interval of the spectral analysis, the PL or CPL index $\alpha$, the CPL peak energy $E_{\rm p}$ when present, the BB observed temperature $kT_{\rm obs}$ and normalization $A_{\rm BB}$, fitted from Sec.~\ref{sec3} . The quantity $\phi_0$, the expansion velocity $\beta$ and the Lorentz factor $\Gamma$, and the effective thermal emitter radius in the laboratory frame $R$ inferred from Sec.~\ref{sec4}.}
\tiny
\begin{tabular}{cccccccccc}
\hline\hline
$\Delta t_a^d$	  	&	Model				&	$\alpha$ &  $E_{\rm p}$		&	$kT_{\rm obs}$			&		$A_{\rm BB}$			& 	$\phi_0$			&	$\beta$				&	$\Gamma$			&		$R$				\\
(s)            			&					&		 &	(keV)		&	(keV)				&	(ph~cm$^{-2}$s$^{-1}$)  	&  	($10^{12}$~cm) 			&					&					&    		($10^{12}$~cm)			\\
\hline
$94$--$100$			&	BB$+$PL				&	$1.349_{-0.036}^{+0.024}$      &&	$2.2_{-1.1}^{+1.1}$  		&	$0.052^{+0.043}_{-0.034}$	&	$0.065^{+0.070}_{-0.064}$	&	$0.38^{+0.19}_{-0.31}$		&	$1.079^{+0.138}_{-0.077}$	&	$0.10^{+0.11}_{-0.10}$		\\
$100$--$110$			&	BB$+$PL				&	$1.293_{-0.031}^{+0.029}$      &&	$2.57^{+0.43}_{-0.50}$		&	$0.206^{+0.083}_{-0.084}$	&	$0.094^{+0.037}_{-0.041}$	&	$0.606^{+0.042}_{-0.049}$	&	$1.257^{+0.057}_{-0.053}$	&	$0.194^{+0.077}_{-0.086}$	\\
$110$--$120$			&	BB$+$PL				&	$1.392_{-0.033}^{+0.028}$      &&	$2.17^{+0.22}_{-0.26}$  	&	$0.62^{+0.14}_{-0.15}$		& 	$0.229^{+0.053}_{-0.062}$	&	$0.852^{+0.035}_{-0.052}$	&	$1.91^{+0.26}_{-0.24}$		&	$0.80^{+0.21}_{-0.25}$		\\
$120$--$130$			&	BB$+$PL				& 	$1.732_{-0.057}^{+0.049}$      &&	$1.10^{+0.14}_{-0.12}$		&	$0.592^{+0.077}_{-0.073}$	&	$0.87^{+0.23}_{-0.20}$		&	$0.957^{+0.014}_{-0.028}$	&	$3.46^{+0.78}_{-0.76}$		&		$5.7^{+1.8}_{-2.3}$		\\
$130$--$140$			&	BB$+$PL				&	$1.82_{-0.14}^{+0.11}$         &&	$0.617^{+0.046}_{-0.043}$    	&	$0.247^{+0.037}_{-0.038}$	& 	$1.79^{+0.30}_{-0.28}$		&	$0.983^{+0.0046}_{-0.0079}$	&	$5.6^{+1.0}_{-1.0}$		&	$19.1^{+4.2}_{-5.6}$		\\
$140$--$150$			&	CPL$+$PL			&  $1.65_{-0.16}^{+0.15}$ & $7.3^{+66.3}_{-4.6}$	&	$0.469^{+0.065}_{-0.064}$ &	$0.102^{+0.028}_{-0.027}$	& 	$1.99^{+0.61}_{-0.61}$		&	$0.919^{+0.054}_{-0.560}$	&	$2.5^{+1.8}_{-1.5}$		&		$9.5^{+4.4}_{-9.5}$		\\
$150$--$160$			&	BB$+$PL				&	$2.40_{-0.34}^{+0.45}$         &&	$0.386^{+0.061}_{-0.061}$     	&	$0.046^{+0.016}_{-0.015}$	& 	$1.97^{+0.71}_{-0.70}$ 		&	$0.935^{+0.048}_{-0.934}$	&	$2.8^{+2.7}_{-1.8}$		&	$10.5^{+5.5}_{-10.5}$		\\
$160$--$180$			&	BB$+$PL				&	$2.15_{-0.34}^{+0.29}$         &&	$0.193^{+0.032}_{-0.030}$    	&	$0.020^{+0.011}_{-0.013}$	& 	$5.2^{+2.3}_{-2.3}$		&	$0.953^{+0.042}_{-0.952}$	&	$3.3^{+7.0}_{-2.3}$		&		$32^{+21}_{-32}$		\\
\hline												
\end{tabular}
\label{tab1}
\end{table*}

{Following Fig.~\ref{fig:funcV} it is possible to infer the expansion velocity $\beta$ (i.e., the velocity in units of the velocity of light $c$). We assume that the black body emitter has spherical symmetry and expands with a constant Lorentz gamma factor. Therefore, the expansion velocity $\beta$ is also constant during the emission. The relations between the comoving time $t_{com}$, the laboratory time $t$, the arrival time $t_a$, and the arrival time $t_a^d$ at the detector \citep[see][]{2001A&A...368..377B,Ruffini2001L107,2002ApJ...581L..19R,Bianco2005a} in this case become:
\begin{align}
t_a^d & = t_a(1+z) = t (1-\beta\cos\vartheta)(1+z)\nonumber \\
& = \Gamma t_{com} (1-\beta\cos\vartheta)(1+z)\ .
\label{times}
\end{align}
We can infer an effective radius $R$ of the black body emitter from: 1) the observed black body temperature $T_\mathrm{obs}$, which comes from the spectral fit of the data; 2) the observed bolometric black body flux $F_\mathrm{bb,obs}$, computed from $T_\mathrm{obs}$ and the normalization of the black body spectral fit; and 3) the cosmological redshift $z$ of the source \citep[see also][]{2012A&A...543A..10I}. We recall that $F_\mathrm{bb,obs}$ by definition is given by:
\begin{equation}
F_\mathrm{bb,obs} = \frac{L}{4\pi D_L(z)^2}\ ,
\label{fbbobs0}
\end{equation}
where $D_L(z)$ is the luminosity distance of the source, which in turn is a function of the cosmological redshift $z$, and $L$ is the source bolometric luminosity (i.e., the total emitted energy per unit time). $L$ is Lorentz invariant, so we can compute it in the co-moving frame of the emitter using the usual black body expression:
\begin{equation}
L=4\pi {R_\mathrm{com}}^2 \sigma {T_\mathrm{com}}^4\ ,
\label{lum}
\end{equation}
where $R_\mathrm{com}$ and $T_\mathrm{com}$ are the comoving radius and the comoving temperature of the emitter, respectively, and $\sigma$ is the Stefan-Boltzmann constant. We recall that $T_\mathrm{com}$ is constant over the entire shell due to our assumption of spherical symmetry. From Eq.~(\ref{fbbobs0}) and Eq.~(\ref{lum}) we then have:
\begin{equation}
F_\mathrm{bb,obs}= \frac{{R_\mathrm{com}}^2 \sigma {T_\mathrm{com}}^4}{D_L(z)^2}\ .
\label{fbbobs1}
\end{equation}
}

{We now need the relation between $T_\mathrm{com}$ and the observed black body temperature $T_\mathrm{obs}$. Considering both the cosmological redshift and the Doppler effect due to the velocity of the emitting surface, we have:
\begin{align}
T_\mathrm{obs} (T_\mathrm{com},z,\Gamma,\cos\vartheta) &= \frac{T_\mathrm{com}}{\left(1+z\right)\Gamma\left(1-\beta\cos\vartheta\right)}\nonumber\\
&= \frac{T_\mathrm{com}\mathcal{D}(\cos\vartheta)}{1+z}\, ,
\label{tdef}
\end{align}
where we have defined the Doppler factor $\mathcal{D}(\cos\vartheta)$ as:
\begin{equation}
\mathcal{D}(\cos\vartheta)\equiv\frac{1}{\Gamma\left(1-\beta\cos\vartheta\right)}\, .
\label{defd}
\end{equation}
Eq.~(\ref{tdef}) gives us the observed black body temperature of the radiation coming from different points of the emitter surface, corresponding to different values of $\cos\vartheta$. However, since the emitter is at a cosmological distance, we are not able to resolve spatially the source with our detectors. Therefore, the temperature that we actually observe corresponds to an average of Eq.~(\ref{tdef}) computed over the emitter surface:
\begin{align}
\nonumber T_\mathrm{obs}(T_\mathrm{com},z,\Gamma)=&\,\displaystyle\frac{1}{1+z}\frac{\int^{1}_{\beta}{\mathcal{D}(\cos\vartheta)T_\mathrm{com}\cos\vartheta d\cos\vartheta}}{\int^{1}_{\beta}{\cos\vartheta d\cos\vartheta}} \\[6pt]
\nonumber=&\, \displaystyle\frac{2}{1+z}\frac{\beta\left(\beta-1\right)+\ln\left(1+\beta\right)}{\Gamma\beta^2\left(1-\beta^2\right)}T_\mathrm{com}\\[6pt]
\label{tobsbb}=&\,\Theta(\beta)\frac{\Gamma}{1+z}T_\mathrm{com}
\end{align}
where we defined
\begin{equation}
\Theta(\beta) \equiv 2\, \frac{\beta\left(\beta-1\right)+\ln\left(1+\beta\right)}{\beta^2}\, ,
\label{ThetaDef}
\end{equation}
we have used the fact that due to relativistic beaming, we observe only a portion of the surface of the emitter defined by:
\begin{equation}
\beta \leq \cos\vartheta \leq 1\, ,
\label{visible}
\end{equation}
and we used the definition of $\Gamma$ given above. Therefore, inverting Eq.~(\ref{tobsbb}), the comoving black body temperature $T_\mathrm{com}$ can be computed from the observed black body temperature $T_\mathrm{obs}$, the source cosmological redshift $z$ and the emitter Lorentz gamma factor in the following way:
\begin{equation}
T_\mathrm{com} (T_\mathrm{obs},z,\Gamma) = \frac{1+z}{\Theta(\beta)\Gamma}T_\mathrm{obs}\, .
\label{tcomdef}
\end{equation}
}

\begin{figure}[ht]
\centering
\includegraphics[width=\hsize,clip]{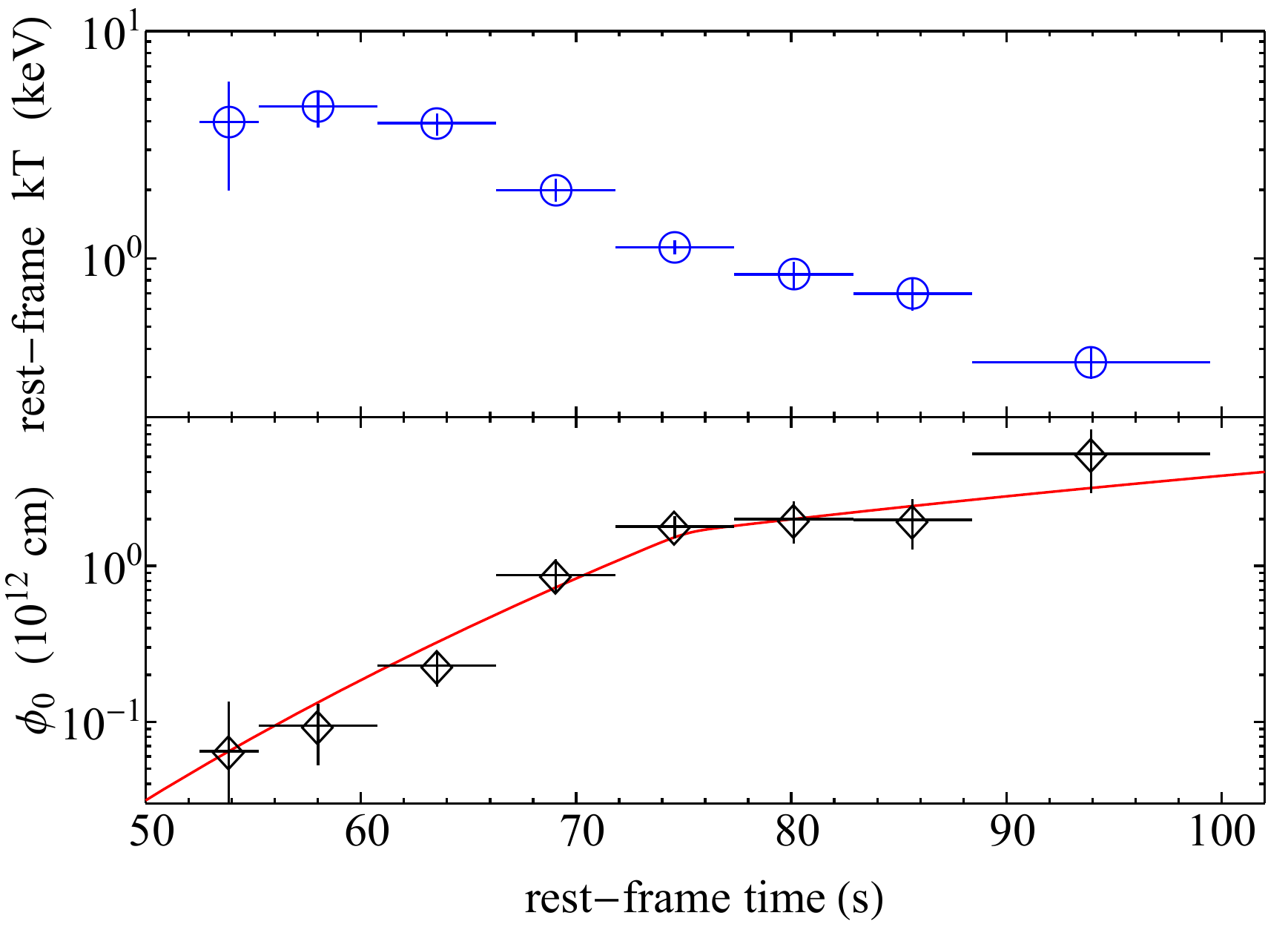}
\caption{The cosmological rest-frame evolution of $kT$ (upper panel) and $\phi_0$ (bottom panel) of the thermal emitter in the hard X-ray flare of GRB 151027A. The $\phi_0$ interpolation (red line) is obtained by using two smoothly joined PL segments.}
\label{fig2}
\end{figure}

{We can now insert Eq.~(\ref{tcomdef}) into Eq.~(\ref{fbbobs1}) to obtain:
\begin{equation}
F_\mathrm{bb,obs} = \frac{{R_\mathrm{com}}^2}{D_L(z)^2} \sigma T_\mathrm{com}^4 = \frac{{R_\mathrm{com}}^2}{D_L(z)^2} \sigma \left[\frac{1+z}{\Theta(\beta)\Gamma}T_\mathrm{obs}\right]^4\, .
\label{fbbobs2}
\end{equation}
Since the radius $R$ of the emitter in the laboratory frame is related to $R_\mathrm{com}$ by:
\begin{equation}
R_\mathrm{com} = \Gamma R\, ,
\label{rcomdef}
\end{equation}
we can insert Eq.~(\ref{rcomdef}) into Eq.~(\ref{fbbobs2}) and obtain:
\begin{equation}
\label{fbbobs} F_\mathrm{bb,obs}=\frac{\left(1+z\right)^4}{\Gamma^2}\left(\frac{R}{D_L(z)}\right)^2\sigma \left[\frac{T_\mathrm{obs}}{\Theta(\beta)}\right]^4\ .
\end{equation}
Solving Eq.~(\ref{fbbobs}) for $R$ we finally obtain the thermal emitter effective radius in the laboratory frame:
\begin{equation}
\label{raggiorel} R=\Theta(\beta)^2\Gamma\frac{D_L(z)}{(1+z)^2}\sqrt{\frac{F_\mathrm{bb,obs}}{\sigma T_\mathrm{obs}^4}}=\Theta(\beta)^2\Gamma \phi_0\ ,
\end{equation}
where we have defined $\phi_0$:
\begin{equation}
\phi_0 \equiv \frac{D_L(z)}{(1+z)^2}\sqrt{\frac{F_\mathrm{bb,obs}}{\sigma T_\mathrm{obs}^4}}\, .
\label{rclass}
\end{equation}
The evolutions of the rest-frame temperature and $\phi_0$ are shown in Fig.~\ref{fig2}. In astronomy the quantity $\phi_0$ is usually identified with the radius of the emitter. However, in relativistic astrophysics this identity cannot be straightforwardly applied, because the estimate of the effective emitter radius $R$ in Eq.~\ref{raggiorel} crucially depends on the knowledge of its expansion velocity $\beta$ (and, correspondingly, of $\Gamma$).
}

{
It must be noted that Eq.~(\ref{raggiorel}) above gives the correct value of $R$ for all values of $0 \leq \beta \leq 1$ by taking all the relativistic transformations properly into account. In the non-relativistic limit ($\beta \rightarrow 0$) we have respectively:
\begin{align}
&\Theta\xrightarrow[\beta\rightarrow 0]{} 1\, , &\Theta^2\xrightarrow[\beta\rightarrow 0]{} 1\, , \\
&T_\mathrm{com}\xrightarrow[\beta\rightarrow 0]{}T_\mathrm{obs}(1+z)\, , &R\xrightarrow[\beta\rightarrow 0]{}\phi_0\, ,
\end{align}
as expected. Analogously, in the ultrarelativistic limit ($\beta \rightarrow 1$) we have:
\begin{align}
&\Theta\xrightarrow[\beta\rightarrow 1]{} 1.39\, , &\Theta^2\xrightarrow[\beta\rightarrow 1]{} 1.92\, , \\
&T_\mathrm{com}\xrightarrow[\beta\rightarrow 1]{}\frac{0.72}{\Gamma}T_\mathrm{obs}(1+z)\, , &R\xrightarrow[\beta\rightarrow 1]{}1.92\Gamma\phi_0\, ,
\end{align}
It must also be noted that the numerical coefficient in Eq.(\ref{raggiorel}) is computed as a function of $\beta$ using Eq.(\ref{ThetaDef}) above, and it is different from the constant values proposed by \citet{2007ApJ...664L...1P} and by \citet{2013MNRAS.432.3237G}.
}

An estimate of the expansion velocity $\beta$ can be deduced from the ratio between the variation of the emitter effective radius $\Delta R$ and the emission duration in laboratory frame $\Delta t$, i.e., 
\begin{equation}
\beta=\frac{\Delta R}{c \Delta t}=\Theta(\beta)^2\Gamma(1-\beta\cos\vartheta)(1+z)\frac{\Delta \phi_0}{c \Delta t_a^d}\, ,
\label{beta1}
\end{equation}
where we have used Eq.~(\ref{raggiorel}) and the relation between $\Delta t$ and $\Delta t_a^d$ given in Eq.~(\ref{times}), we have used the definition of $\Gamma$ given above and $\vartheta$ is the displacement angle of the considered photon emission point on the surface from the line of sight. In the following we consider only the case $\cos\vartheta=1$. In this case, using Eq.(\ref{ThetaDef}), Eq.(\ref{beta1}) assumes the form presented in Fig.~\ref{fig:funcV}. It allows to estimate the expansion velocity $\beta$ of the emitter using only the observed black body flux, temperature, photon arrival time, and cosmological redshift, assuming uncollimated emission and considering only the radiation coming from the line of sight. We can explain the observed black body emission in GRB 151027A without introducing the ``reborn fireball'' scenario \citep[see][]{2007MNRAS.382L..72G,2017A&A...598A..23N}.
\begin{figure}[ht]
\centering
\includegraphics[width=\hsize,clip]{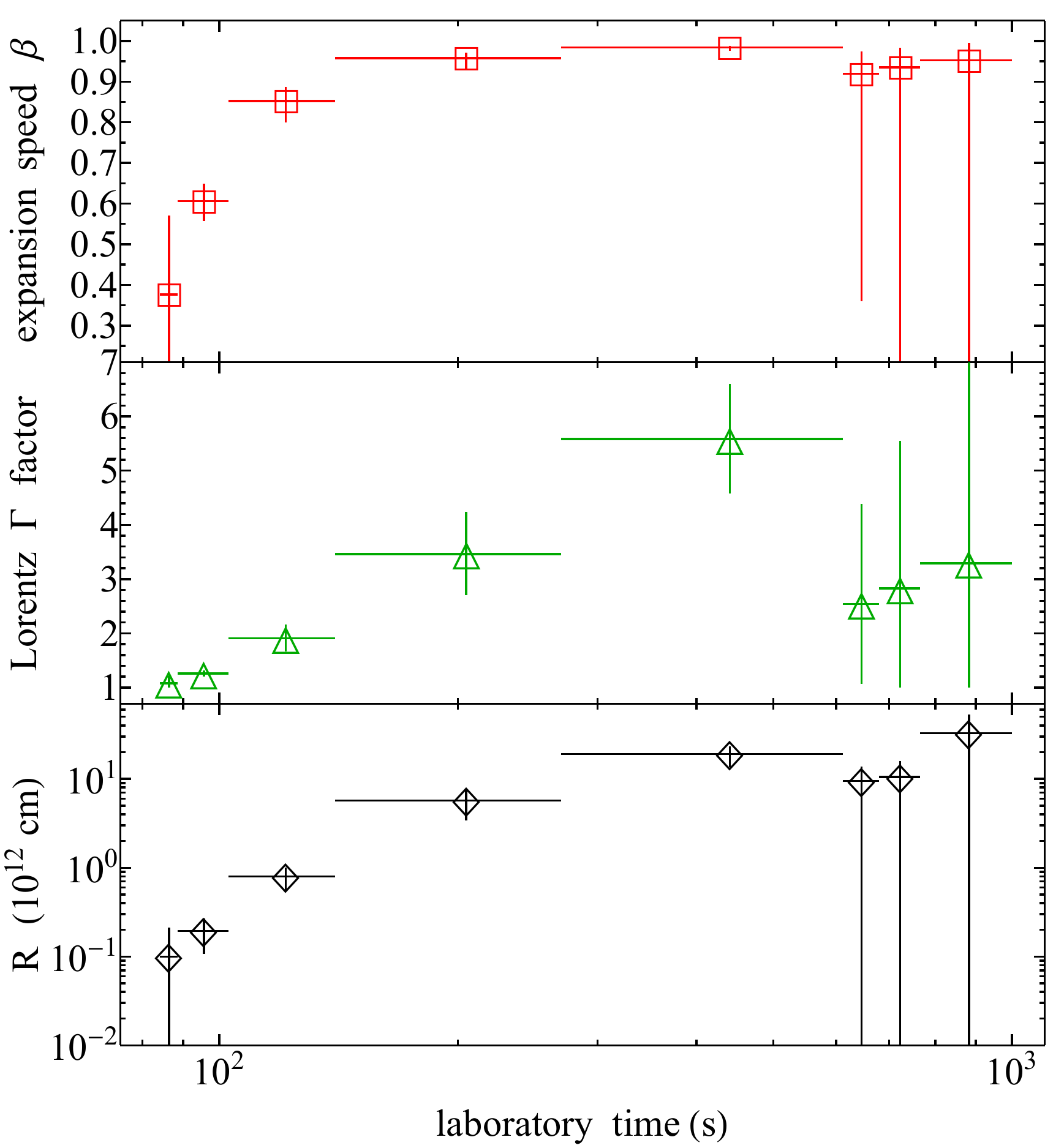}
\caption{The evolution in the laboratory frame of $\beta$, $\Gamma$ and $R$ of the thermal emitter from the time intervals in Tab.~\ref{tab1}.}
\label{fig3}
\end{figure}

To infer $\beta$, we fit the evolution of $\phi_0$ (see Fig.~\ref{fig2} and Tab.~\ref{tab1}) by using two smoothly joined PL segments. It allows us to estimate the ratio $\Delta \phi_0/(c \Delta t_a^d)$ in Eq.~(\ref{beta1}) and, therefore, the values of $\beta$ and $\Gamma$ assuming that they are constant in each time interval (see Fig.~\ref{fig3}, upper and middle panels). Consequently, we can estimate the evolution of the radius $R$ of the emitter in the laboratory frame by taking into account the relativistic transformations described in Eqs.~(\ref{times}), (\ref{raggiorel}), (\ref{rclass}); see lower panel of Fig.~\ref{fig3}. The results are summarized also in Tab.~\ref{tab1}.

\section{On the nature of the hard X-ray flare and the soft X-ray flare}\label{sec5}

\begin{figure*}
\centering
(a)\hfill$t=0$~s\hfill\hfill(b)\hfill$t=56.7$~s\hfill\hfill(c)\hfill$t=236.8$~s\hfill\null\\
\includegraphics[width=\hsize,clip]{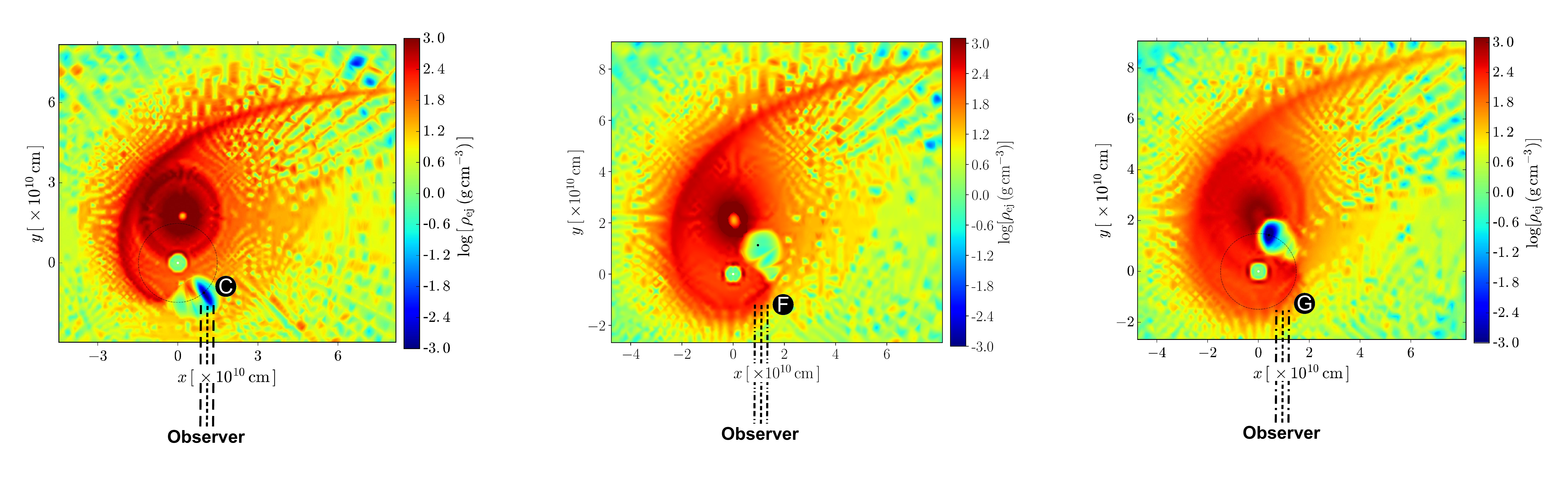}
\caption{Three snapshots of the density distribution of the SN ejecta in the equatorial plane of the progenitor binary system. The time $t=0$ indicates the instant when the NS companion reaches, by accretion, the critical mass and leads to the formation of a BH (black dot). As evidenced in panel (a), the location of the black hole formation is widely separated from the central position represented by SN explosion, it is actually located in the white conical region in Fig.~\ref{fig:cc}. The binary parameters of this simulations are: the NS companion has an initial mass of $2.0~M_\odot$; the CO$_{\rm core}$, obtained from a progenitor with ZAMS mass $M_{\rm ZAMS}=30~M_\odot$, leads to a total ejecta mass $7.94~M_\odot$ and to a $1.5~M_\odot$ $\nu$NS (white dot), the orbital period is $P\approx 5$~min, i.e. a binary separation $a\approx 1.5\times 10^{10}$~cm.}
\label{fig:Carlo2}
\end{figure*}

Following the procedure described in section~10 of \citet{2018ApJ...852...53R}, we interpret the thermal emission observed in the hard X-ray flare as the observational feature arising from the early interaction between the expanding SN ejecta and the $e^+e^-$ plasma. In order to test the consistency of this model with the data, we have performed a series of numerical simulations, whose details we summarize as follows.

a) Our treatment of the problem is based on an implementation of the one-dimensional relativistic hydrodynamical (RHD) module included in the PLUTO code\footnote{http://plutocode.ph.unito.it/} \citep{PLUTO}. In the spherically symmetric case considered, only the radial coordinate is used, and consequently the code integrates a system of partial differential equations in only two coordinates: the radius and the time. This permits the study of the evolution of the plasma along one selected radial direction at a time. The aforementioned equations are those of an ideal relativistic fluid, which can be written as follows:
\begin{align}
&\frac{\partial(\rho \Gamma)}{\partial t} +\nabla.\left(\rho\Gamma
\mathbf{v}\right)=0, \label{consmass}\\
&\frac{\partial m_r}{\partial t} +\nabla.\left(m_r
\mathbf{v}\right)+\frac{\partial p}{\partial r}=0,\label{consmomentum}\\
&\frac{\partial \mathcal{E}}{\partial t}+ \nabla .\,\left(\mathbf{m}-\rho\Gamma\mathbf{v}\right)=0,\label{consenergy}
\end{align}
where $\rho$ and $p$ are the comoving fluid density and pressure, $\mathbf{v}$ is the coordinate velocity in natural units ($c=1$), $\Gamma=(1-\mathbf{v}^2)^{-\frac{1}{2}}$ is the Lorentz gamma factor, $\mathbf{m}=h\Gamma^2\mathbf{v}$ is the fluid momentum, $m_r$ its radial component, $\mathcal{E}$ is the internal energy density measured in the comoving frame, and $h$ is the comoving enthalpy density which is defined by $h=\rho+\epsilon+p$. We define $\mathcal{E}$ as follows:
\begin{equation}
\mathcal{E}=h\Gamma^2-p-\rho\Gamma.
\end{equation}
The first two terms on the right-hand side of this equation coincide with the $T^{00}$ component of the fluid energy-momentum, and the last one is the mass density in the  laboratory frame.

\begin{figure}[ht]
\centering
\includegraphics[width=1\hsize,clip]{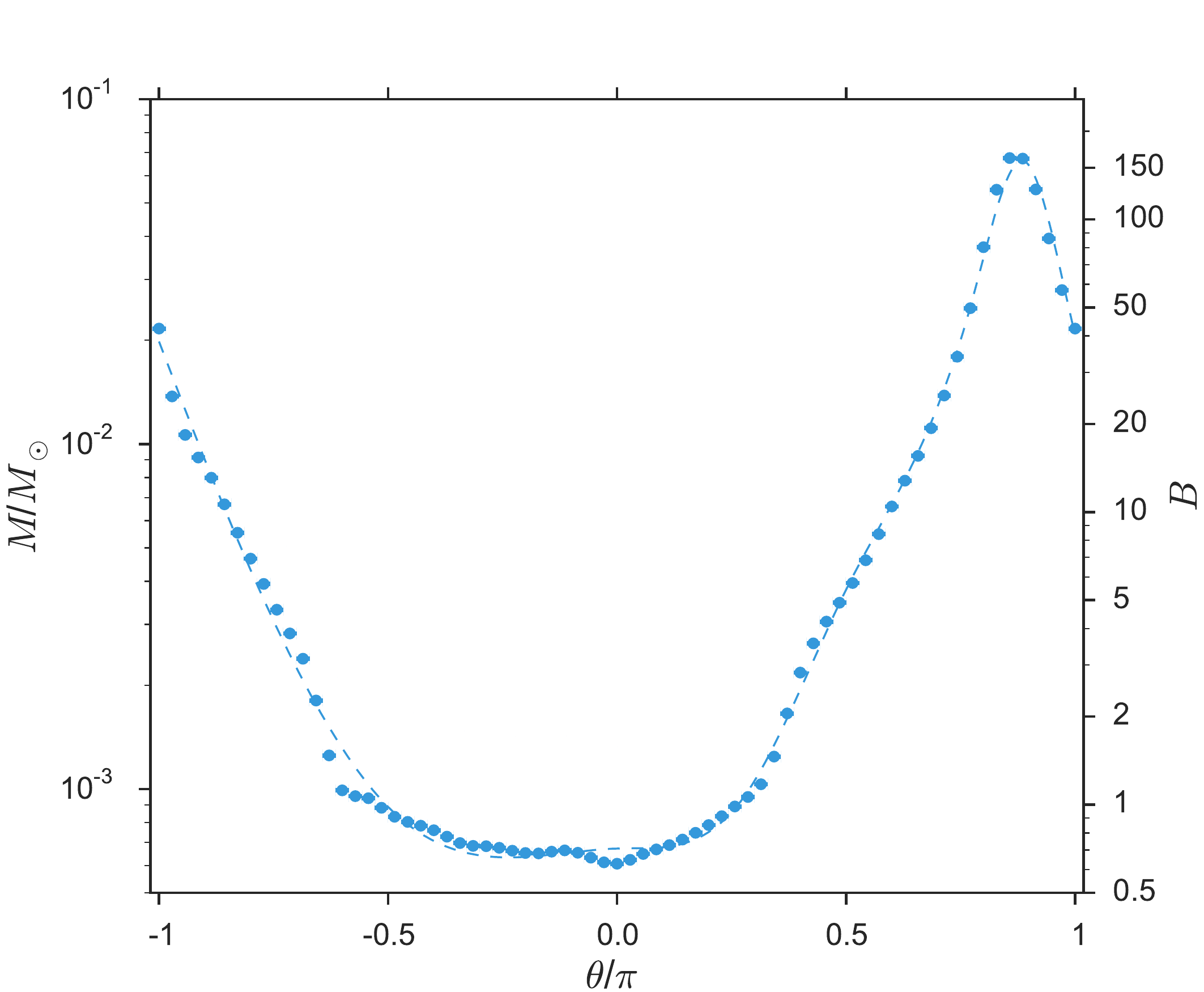}
\caption{The SN ejecta mass enclosed within a cone of 5 degrees of semi-aperture angle and vertex centered on the SN and positioned to an angle $\theta$, measured counterclockwise, with respect to the line of sight (which passes through the $\nu$NS and BH at the moment of its formation; see Conclusions). The binary parameters of this simulations are: the NS has an initial mass of $2.0~M_\odot$; the CO$_{\rm core}$ obtained from a progenitor with ZAMS mass $M_{\rm ZAMS}=30~M_\odot$, leads to a total ejecta mass $7.94~M_\odot$, the orbital period is $P\approx 5$~min, i.e. a binary separation $a\approx 1.5\times 10^{10}$~cm. The right-side vertical axis gives, as an example, the corresponding value of the baryon load $B$ assuming a plasma energy of $E_{e^+e^-}=1\times10^{53}$~erg. It is appropriate to mention that the above values of the baryon load are computed using an averaging procedure which is performed centered on the SN explosion, and which produces larger values than the one centered around the BH with specific value of baryon load $B \sim 1.9 \times  10^{-3}$, see Fig.~\ref{fig:Carlo2}a.}
\label{fig:model1}
\end{figure}

Under the conditions discussed in \citet{2018ApJ...852...53R}, the plasma satisfies the equation of state of an ideal relativistic gas, which can be expressed in terms of its enthalpy as:
\begin{equation}
h=\rho+\frac{\gamma p}{\gamma-1},
\label{eq:eos}
\end{equation}
with $\gamma=4/3$. Imposing this equation of state closes and defines completely the system of equations, leaving as the only remaining freedom the choice of the matter density profile and the boundary conditions. To compute the evolution of these quantities in the chosen setup, the code uses the HLLC Riemann solver for relativistic fluids \citep[see][]{PLUTO}. The time evolution is performed by means of a second-order Runge-Kutta integration, and a second-order total variation diminishing scheme is used for the spatial interpolation. An adaptive mesh refinement algorithm is implemented as well, provided by the CHOMBO library \citep{CHOMBO}. We turn now to the determination of the SN ejecta.

b) {The initially ultrarelativistic $e^+e^-$ plasma expands through the SN ejecta matter slowing down to mildly relativistic velocities}. The SN density and velocity profiles are taken from the 3D SPH simulation of the SN ejecta expansion under the influence of the $\nu$NS and the NS companion gravitational field. In our simulations we include the NS orbital motion and the NS gravitational-mass changes due to the accretion process modeled with the Bondi-Hoyle formalism \citep[see][for more details]{2016ApJ...833..107B}. We set the SN ejecta initial conditions adopting an homologous velocity distribution in free expansion and the SN matter was modeled with 16 million point-like particles. Each SN layer is initially populated following a power-law density profile of the CO$_{\rm core}$ as obtained from low-metallicity progenitors evolved with the Kepler stellar evolution code \citep{2002RvMP...74.1015W}. We take here as reference model the simulation of an initial binary system formed by a $2.0\,M_\odot$ NS and a CO$_{\rm core}$ produced by a $M_{\rm ZAMS} = 30\, M_\odot$ progenitor. This leads to a total ejecta with mass $7.94\, M_\odot$ and a $\nu$NS of $1.5\,M_\odot$. The orbital period of the binary is $P\approx 5$~min, i.e. a binary separation $a \approx 1.5 \times 10^{10}$~cm. The density profile exhibiting the evolution of the SN ejecta and the companion star is shown in Fig.~\ref{fig:Carlo2}. Fig.~\ref{fig:model1} shows the SN ejecta mass enclosed within a cone of 5 degrees of semi-aperture angle, and with vertex at the position of the BH at the moment of its formation. The cone axis stands along the $\theta$ direction measured counterclockwise with respect to the line of sight. We simulate the interaction of the $e^+e^-$ plasma with such ejecta from a radius $\approx10^{10}$~cm all the way to $\approx10^{12}$~cm where transparency is reached. We have recently run new 3D SPH simulations of this process in \citet{2018arXiv180304356B} using the SNSPH code \citep{2006ApJ...643..292F}. These new simulations have allowed a wide exploration of the binary parameter space and have confirmed the results and the physical picture presented in \citet{2016ApJ...833..107B}. On the basis of these new simulations we have determined the value of the baryon loads both for the hard X-ray flares and the soft X-ray flares.

\begin{figure}[ht]
\centering
\includegraphics[width=\hsize,clip]{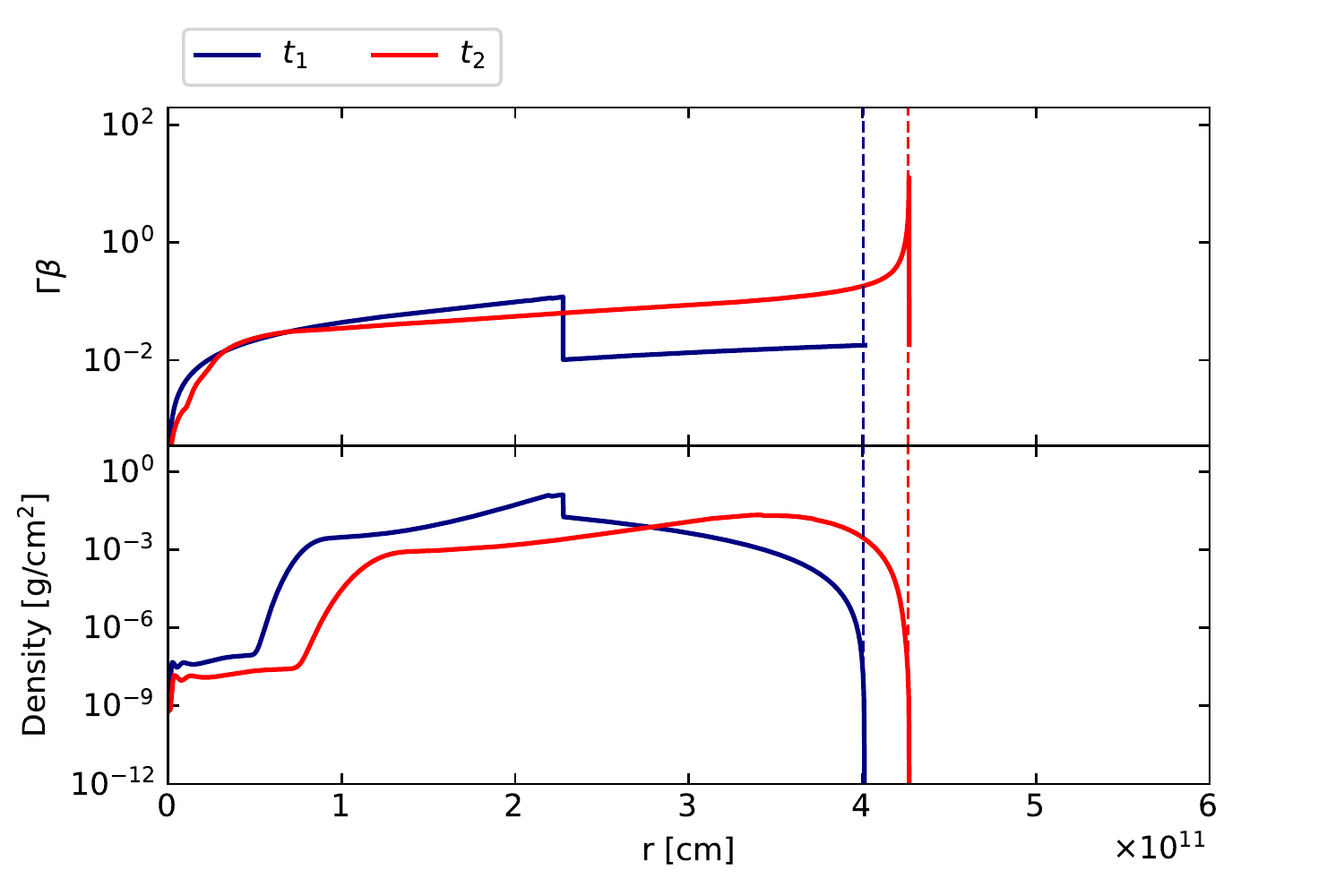}
\caption{Numerical simulation of the hard X-ray flare. We set a total energy of the plasma $E_{\gamma}=3.28\times10^{52}$~erg and a baryon load $B=79$, corresponding to a baryonic mass of $M_B=1.45$~M$_\odot$. \textbf{Above:} Distribution of the velocity inside the SN ejecta at the two fixed values of the laboratory time $t_1$ (before the plasma reaches the external surface of the ejecta) and $t_2$ (the moment at which the plasma, after having crossed the entire SN ejecta, reaches the external surface). We plotted the quantity $\Gamma\beta$, recalling that we have $\Gamma\beta \sim \beta$ when $\beta < 1$ and $\Gamma\beta \sim \Gamma$ when $\beta \sim 1$. \textbf{Below:} Corresponding distribution of the mass density of the SN ejecta in the laboratory frame $\rho_{lab}$. We obtain a radius of the transparency $R_{ph}=4.26\times10^{11}$~cm, a Lorentz factor at transparency $\Gamma=2.86$ and an arrival time of the corresponding radiation in the cosmological rest frame $t_a=56.7$~s.}
\label{fig:gamma_ray_flare}
\end{figure}

c) For the simulation of the hard X-ray flare we set a total energy of the plasma equal to that of the hard X-ray flare, i.e., $E_{\gamma}=3.28\times10^{52}$~erg, and a baryon load $B=79$, corresponding to a baryonic mass of $M_B=1.45~M_\odot$. We obtain a radius of the transparency $R_{ph}=4.26\times10^{11}$~cm, a Lorentz factor at transparency $\Gamma=2.86$ and an arrival time of the corresponding radiation in the cosmological rest frame $t_a=56.7$~s (see Fig.~\ref{fig:gamma_ray_flare}). This time is in agreement with the starting time of the hard X-ray flare in the source rest-frame (see Sec.~\ref{sec2}).

\begin{figure}[ht]
\centering
\includegraphics[width=\hsize,clip]{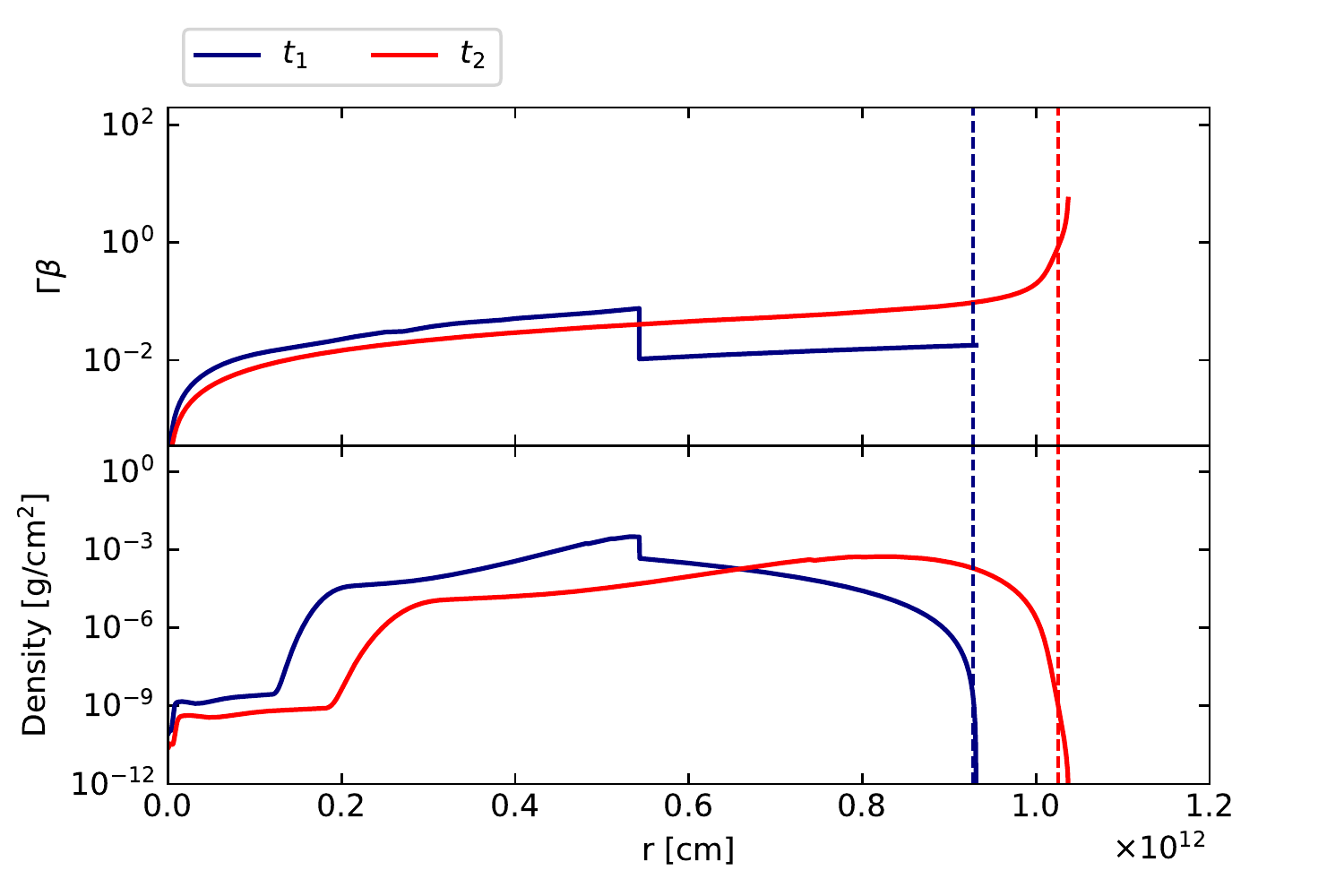}
\caption{Numerical simulation of the soft X-ray flare. We set a total energy of the plasma $E_X=4.39\times 10^{51}$~erg and a baryon load $B=207$, corresponding to a baryonic mass of $M_B=0.51~M_\odot$. The plotted quantities are the same as in Fig.~\ref{fig:gamma_ray_flare}. We obtain a radius of the transparency $R_{ph}=1.01\times10^{12}$~cm, a Lorentz factor at transparency $\Gamma=1.15$ and an arrival time of the corresponding radiation in the cosmological rest frame $t_a=236.8$~s.}
\label{fig:x_ray_flare}
\end{figure}

For the simulation of the soft X-ray flare we set the energy $E_X=4.39\times 10^{51}$~erg as the total energy of the plasma and a baryon load $B=207$, which corresponds to a baryonic mass of $M_B=0.51$~M$_\odot$, we obtain a radius of the transparency $R_{ph}=1.01\times10^{12}$~cm, a Lorentz gamma factor at transparency $\Gamma=1.15$ and an arrival time of the corresponding radiation in the cosmological rest frame $t_a=236.8$~s (see Fig.~\ref{fig:x_ray_flare}). This time is in agreement with the above time $t_p$ at which the soft X-ray flare peaks in the rest frame.

\begin{table*}
\centering
\small\addtolength{\tabcolsep}{-5pt}
\begin{tabular}{lccccccc}
\hline\hline
 &Name& Radius(cm) & $\Gamma$ & Baryon load &$t_{start}(s)$&duration(s)& Spectrum\\
\hline
                  &  First spike(P-GRB)  &  $\sim10^{13}$  & $\sim 10^{2}-10^{3}$&   $\sim 10^{-4}-10^{-2}$ & $\sim T_0$& $ \sim 1$& CPL+BB\\
        UPE \Bigg\{       & First spike(Rest) &  $\sim 10^{15}-10^{17}$  & $\sim 10^{2}-10^{3}$&   $ \sim10^{-4}-10^{-2}$& $\sim T_0 +1$& $\sim 5$ & Band\\
                
                  &Second spike &  $\sim 10^{15}-10^{17}$  & $\gtrsim 10^{3}$&   $\lesssim  10^{-4}$&$\sim T_0 +15$&$\sim 5$ &Band\\
\hline
 &hard X-ray flare& $\sim 10^{11}-10^{12}$  & $\lesssim 10$ &   $\sim 10^{2}$ &$\sim T_0 +50$&$\sim 10^2$&PL+BB\\
 &soft X-ray flare& $\sim 10^{12}-10^{13}$  & $\lesssim 4$ &   $\sim 10^{3}$&$\sim T_0 +10^2$&$\sim 150$&PL(+BB) \\ 
\hline
& Late Afterglow& $\gtrsim 10^{13}$  & $\lesssim 2$ &   $-$&$\gtrsim T_0 +10^2$&$\gtrsim 10^6 $&PL \\ 
\hline
 & SN optical emission& $\sim 10^{15}$  & $\sim 1$ &   $-$&$\sim T_0 +10^6$&$\gtrsim 10^6 $&PL \\ 
 \hline
 &GeV emission & $-$  & $-$ &   $-$&$\sim T_0 +1$&$\sim 10^4 $&PL \\ 

\hline \hline
\end{tabular}
\caption{{\textit{\textbf{Parameters of sequence of astrophysical processes characterizing the BdHNe}}: The columns list, respectively, the name of each process, the radius of transparency, the Lorentz Gamma factor ($\Gamma$) and the baryon load, starting time of the process, the duration and finally the best-fit model of the spectrum. $T_0$ is the \textit{Fermi}-GBM trigger time.}}
\label{Tab1}
\end{table*}

\section{Summary, Discussion and Conclusions}\label{sec6}

\subsection{Summary}

{
It is by now clear that seven different subclass of GRBs with different progenitors exist \citep{2016ApJ...832..136R}. Each GRB subclass is itself composed of different episodes each one characterized by specific observational data which make their firm identification possible \citep[see e.g.][and references therein]{2018ApJ...852...53R}. We here evidence how, within the BdHN subclass, a further differentiation follows by selecting special viewing angles. We have applied our recent treatment \citep{2018ApJ...852...53R} to the UPE phase and the hard X-ray flare using as a prototype the specific case of GRB 151027A in view of the excellent available data.}

{
We recall three results:
\begin{enumerate}
\item 
We have confirmed the ultrarelativistic nature of the UPE which appear to be composed by a double spike; see Figs.~\ref{fig1a}(a) and \ref{fig0a}(b). This double spike structure appears to be also present in other systems such as GRB 140206A and GRB 160509A (Ruffini, et al., in preparation). From the analysis of the P-GRB of the first spike we have derived an ultra-relativistic Lorentz factor $\Gamma_0 = 503\pm 76$, a baryon load $B=(1.92\pm0.35)\times10^{-3}$, and a structure in the CBM with density $(7.46\pm1.2)$~cm$^{-3}$ extending to dimensions of $10^{16}$~cm; see Fig.~\ref{fig0a}(d). The second spike of energy $E_{\rm iso,2} = (4.99\pm 0.60)\times 10^{51}$~erg, following by $9$~s in the cosmological rest frame the first spike of energy $E_{\rm iso,1} = (7.26\pm 0.36)\times 10^{51}$~erg, see Fig.~\ref{fig1a}(b) and (c), appears to be featureless. We are currently examining the possibility that the nature of these two spikes and their morphology be directly connected to the formation process of the BH.
\item
A double spikes appears to occur also in the FPA phase (see Fig.~\ref{fig1b}(a)): the first component is the hard X-ray flare and the second is the soft X-ray flare. The energy of the hard X-ray flare is $E_{\gamma}=(3.28\pm0.13)\times10^{52}$~erg (Fig.~\ref{fig1b}) and the one of the soft X-ray flare is $E_X=(4.4 \pm 2.9)\times 10^{51}$~erg (Fig.~\ref{fig1c}). We have analyzed both flares by our usual approach of the hydrodynamical equations describing the interaction of the $e^+e^-$ plasma with the SN ejecta: see Fig.~\ref{fig:gamma_ray_flare} for the hard X-ray flare and Fig.~\ref{fig:x_ray_flare} for the soft X-ray flare.  The baryon load of the two flares are different, $B=79$ for the hard X-ray flare and $B=207$ for the soft X-ray flare. This is visualized in Fig.~\ref{fig:03b} as well as in our three-dimensional simulations; see the three snapshots shown in Fig.~\ref{fig:Carlo2}. Both the hard X-ray flare and the soft X-ray flare show mildly-relativistic regime, already observed in \citet{2018ApJ...852...53R}, namely a Lorentz factor at transparency of $\Gamma\sim 5$ for the hard X-ray flare and a Lorentz factor of $\Gamma\sim 2$ for the soft X-ray flare.
\item
We have studied the ETE associated to the hard X-ray flare: we have measured its expansion velocity derived from the relativistic treatment described in Sec.~\ref{sec4}, following the formula in Fig.~\ref{fig:funcV} \citep[see also][]{2018ApJ...852...53R}. We have identified the transition from a SN, with an initial computed velocity of $0.38~c$, to an HN, with a computed velocity of $0.98~c$; see Fig.~\ref{fig3} and Tab.~\ref{tab1}. These results are in good agreement with observations of both SNe and HNe \citep[see e.g.~Table 3 and Fig.~20 in][]{2015MNRAS.452.3869N}.
\end{enumerate}
}

The above observational analysis, as already presented in \citet{Pisani2013,2016ApJ...833..159P}, set the ensemble of the data that any viable model of GRBs has to conform. In the last thirty years the enormous number of high quality data obtained e.g. by Beppo-SAX, Swift, Agile and Fermi, further extended by specific optical, radio and ultrahigh-energy data, offered the possibility to test the viable models which conform to these data. We have shown that the BdHN model can explain the above observational features.

\subsection{Discussion}

{
\begin{enumerate}
\item 
Thanks to adopting the BdHN approach we have discovered the existence of four different process: a double feature in the UPE phase, the Hard X-ray Flares and the Soft X-ray Flares and  the ETE phase. Each one of these processes is generated by a different $e^+ e^-$ injection occurring in a different baryon load media. By using the binary nature of the progenitor system in BDHN, especially the presence of an incipient SN and a companion NS, together with an appropriate theoretical treatment and an  ample program of numerical simulations \citep{2018arXiv180304356B}, we have been able to determine the nature of these  processes. Clear observational predictions have followed including, the major one, the coincidence of the numerical value of the velocity of expansion at the end of the ETE phase with the observed expansion velocity of the HN, confirmed in additional BdHN and being currently observationally addressed in additional cases. A clear temporal sequence in the occurrence of these process as well as the specific sequence in the values of the Lorentz a Gamma factors has been established. 
\item 
For the first time there of rotation of the binary system, of the order of 400 seconds, has been  essential in order to untangle the sequence of events discovered and explained in this article, recognizing their a-causal nature and their modulation by the rotation of the progenitor binary system. 
\item 
The above different processes, including the double spiky structure of the UPE phase, the Hard and Soft X-Ray Flares, and the ETE  phase are actually different appearances of the same physical process, the Black Hole formation as seen from different viewing angles due to the rotation of the SN ejecta in the binary system (see Fig.~\ref{fig:Carlo2}) and the consequent angular dependence of the baryon load (see Fig.~\ref{fig:model1}). 
\end{enumerate}
}

\subsection{Conclusions}

{
\begin{enumerate}
\item A clear prediction which will soon be submitted to scrutiny, following from our paper, is that of  all the BdHNe occurring with line of sight in the orbital plane of the binary only a fraction of approximately $10\%$ are actually detectable. They correspond to the sources whose ultra-relativistic emission lies within the allowed cone of $\sim 10^\circ$ of low baryon contamination (see Fig.~\ref{fig:cc} and  Fig.~\ref{fig:model1}). They are the only ones able to trigger the gamma-ray instruments (e.g.~the Fermi/GBM or Swift/BAT detectors). The remaining $90$\% will not be detectable by the current satellites and will need possibly a new mission operating in soft X-rays \citep[like, e.g., THESEUS, see][]{2017arXiv171004638A}.
\item The $E_{\rm iso}$, traditionally defined using an underlying assumption of isotropy of the BH emission, has to be modified by considering an anisotropic emission process. A total energy $E_\mathrm{tot}$, summing the energies of the UPE, of the hard X-ray flare, of the ETE, and of the soft X-ray flare, has to be considered for sources seen in the equatorial plane. It is not surprising that the energy of the hard X-ray flare in GRB 151027A is larger than the one of the UPE, pointing to an anisotropic emission from the BH.
\item When the inclination of the viewing angle is less that $60^\circ$ from the normal to the plane of the binary system, the GeV radiation becomes detectable and its energy, which has been related to the BH rotational energy, will need to be taken into account \citep{2018arXiv180305476R}.
\end{enumerate}
}

\acknowledgments

We acknowledge the referee comments which have significantly helped us in formulating a clearer, logically motivated and well balanced presentation of our results.

\software{PLUTO \citep{PLUTO}, CHOMBO \citep{CHOMBO}, SNSPH \citep{2006ApJ...643..292F}}

\end{document}